\documentclass[a4paper,11pt]{article}

\pdfoutput=1
\usepackage{jheppub}
\usepackage{mathrsfs}
\usepackage{cancel}
\usepackage{graphicx}
\usepackage{subcaption}

\newcommand*{\vv}[1]{\vec{\mkern0mu#1}}

\definecolor{green}{rgb}{0.1,0.8,0.2}

\title{Chern-Simons propagators in AdS$_3$}

\author[a]{Jyotirmoy Bhattacharya}
\affiliation[a]{Department of Physics, \\
Indian Institute of Technology Kharagpur, Kharagpur 721302, India.}

\author[a]{\hspace{-0.1cm}, Anurag Guria}

\author[b]{\hspace{-0.1cm}, Shiroman Prakash}
\affiliation[b]{Department of Physics and Computer Science, \\
Dayalbagh Educational Institute, Agra 282005, India.}

\author[a]{\hspace{-0.1cm}, Aditya Sharma}

\author[c]{and Tarun Sharma}
\affiliation[c]{Department of Physics, \\
Indian Institute of Technology Delhi, Hauz Khas, New Delhi 110016, India.}

\emailAdd{jyoti@phy.iitkgp.ac.in, anuragguria88@gmail.com, sprakash@dei.ac.in,
adsharma.d1d4@gmail.com, tks@physics.iitd.ac.in}

\abstract{ We introduce parity-odd spin-1 harmonic functions in AdS$_3$ and study their properties. We demonstrate that such parity-odd harmonics are related to their parity-even counterparts through the action of a `{\it Chern-Simons operator}\hspace{0.08cm}', which we present as a novelty in this paper. This relation leads to the construction of simultaneous eigen-functions of the Laplacian and the Chern-Simons operators. Subsequently, these harmonic functions are employed to construct propagators in pure abelian Chern-Simons theory as well as Maxwell-Chern-Simons theory in a covariant gauge. We demonstrate the consistency of the Chern-Simons propagator with the expected two-point function of the boundary currents. Our results are built upon the embedding formalism, which we modify suitably to incorporate parity-odd structures. This formalism also readily helps us write down parity odd structures for the propagators of higher-spin fields. Finally, we construct a split representation for the parity-odd harmonic functions, which may be useful to compute Witten diagrams with loops. Our results are expected to be useful in perturbative studies of parity violating QFTs on AdS$_3$.}


\begin{document}


%
\maketitle
\flushbottom
%
\section{Introduction}
%
%
%
Over the last couple of decades, the importance of the AdS/CFT correspondence \cite{Maldacena:1997re, Witten:1998qj} has inspired extensive research on quantum field theories in AdS spacetimes. While in the original correspondence we have dynamical gravity in AdS, recently there has also been considerable interest in studying quantum fields on fixed AdS backgrounds \cite{Aharony:2015zea, Carmi:2018qzm, Ankur:2023lum, Carmi:2019ocp, Carmi:2021dsn, Carmi:2024tzj, Moga:2025gdy, Banados:2022nhj}. In such a scenario, the usual holographic dictionary remains well defined allowing us to compute correlation functions of boundary CFT operators using bulk fields. The boundary CFT does not have an energy-momentum tensor in the usual holographic sense, but its covariance under conformal symmetry remains preserved. In this way QFTs on fixed AdS background can serve as a theoretical laboratory to study various boundary CFTs, helping us to develop a deeper understanding of conformal correlators.

One of our central interest in this paper are the propagators of various fields in AdS, with primary focus on parity violating theories. Several useful tools have been developed in the literature to study AdS propagators. Among them, a powerful framework suitable for studying the bulk fields and boundary conformal correlations simultaneously, is the `embedding formalism' 
\cite{Mikhailov:2002bp, Biswas:2002nk, Costa:2011mg, Costa:2011dw, Penedones:2010ue, Costa:2014kfa, Sleight:2017krf, Parisini:2023nbd}. This formalism exploits the fact that isometry of Euclidean AdS$_{d+1}$, identified with Euclidean conformal algebra of boundary CFT$_d$, is identical to the Lorentz algebra of $d+2$ dimensional Minkowski space $\mathbb M^{d+2}$. In fact, here AdS$_{d+1}$ is considered to be a hyperbolic hypersurface within $\mathbb M^{d+2}$, and bulk propagators, as well as boundary correlators represented by an `embedding polynomial' in $\mathbb M^{d+2}$, with a clear prescription to recover the former from the latter. Constraints of Lorentz invariance imposed on these embedding polynomials restricts the form of the propagators and correlators significantly. Besides, this formalism also helps us identify the right basis of AdS harmonic functions, which can be used to expand the AdS propagators. This spectral representation of the AdS propagators is the analogue of Fourier transform in flat space. It enables us to write down a set of Feynman rules for Witten diagrams, which can be used to perturbatively compute correlation functions in QFTs on AdS. The main advantage of this formalism is that it keeps the entire isometry of AdS manifest which can provide a certain advantage in QFT computations. However, most of this formalism developed so far are applicable for bulk theories which are parity invariant. One of our main objectives in this paper is to generalize the embedding formalism for AdS$_3$ to include parity violating bulk theories, such as theories which includes a Chern-Simons term for spin-1 fields.

Chern-Simons terms are quite common in supergravity theories pertaining to specific examples of AdS/CFT. In fact, the implication of such terms for correlations of dual boundary currents have been discussed by several authors \cite{Witten:1998qj, Freedman:1998tz, Jensen:2010em}. However, these early references do not discuss the propagators and Witten diagrams in such Chern-Simons theories explicitly. In this paper, we develop a detailed understanding of the parity-odd structures in AdS$_3$, which enables us to write down the Chern-Simons propagator explicitly, including a spectral representation for such propagators.

Beside supergravity, QFTs on fixed background AdS involving Chern-Simons terms are also very interesting to explore independently. Pure Chern-Simons theory on AdS is topological, and just like any other manifold with a boundary, there is a boundary current with a well known OPE. As an application of our parity-odd generalizations, we have computed the bulk-to-bulk propagator for the pure abelian Chern-Simons theory. Following the rules of the holographic dictionary, this propagator leads to a boundary current two-point function which is consistent with the expectations of the current-current OPE. As an application of our framework, we also work out the bulk propagators in
massive abelian Chern-Simons theory as well as Maxwell-Chern-Simons theory.

Our Chern-Simons propagator on AdS$_3$, along with its spectral representation, paves the way for performing perturbative computations in Chern-Simons QFTs with interactive spin-1 fields, including their non-Abelian counterparts. Developing a comprehensive knowledge of the CFT data on the boundary of interacting Chern-Simons theories on AdS$_3$ has been the main motivation of our work. Also, large-N Chern-Simons theories with vector matter couplings are known to exhibit non-susy dualities between the bosonic and fermionic versions \cite{Giombi:2011kc, Aharony:2011jz, Aharony:2012nh, Aharony:2012ns, Jain:2013py, Aharony:2018pjn}. These dualities generalize the well known Level-rank duality in non-Abelian Chern-Simons theories \cite{Jain:2013py}. It would be interesting to explore whether such dualities continue to exist in AdS, and if yes, what are its implications for corresponding boundary CFTs. In this paper, we present the initial steps towards answering these intriguing questions.

In our analysis of the Chern-Simons propagators, we have relied heavily on their spectral decomposition in terms of the spin-1 AdS$_3$ harmonic functions. By definition, the spin-1 harmonic functions are eigen-functions of the Laplacian operator. Such spin-1 harmonic functions were previously known only for the parity-even case, which is not adequate for the study of Chern-Simons theories. In order to circumvent this problem, we have unraveled a complete basis of parity-odd divergenceless spin-1 harmonic functions and worked out the associated mathematics. In this context, one of the novelties in our paper, is the introduction of a Chern-Simons operator, which maps divergenceless parity-even harmonic functions to the parity odd ones and vice-versa. This operator constitutes a powerful tool through which we are able to efficiently convert the existing results for parity-even harmonics into those for parity-odd ones. We also identify
the right operator on the embedding space $\mathbb M^{4}$, which inherits the role of this Chern-Simons operator while acting on embedding polynomials. This operator adds significantly to the analytical power of the embedding formalism while dealing with parity-violating theories.

Since the Chern-Simons operator interchanges the divergenceless parity-odd and even spin-1 harmonics,
it was possible to obtain a specific linear combination of them which was a simultaneous eigen-function of the Laplacian as well as the Chern-Simons operators.
Such common eigen-functions exist because the Chern-Simons operator squares to the Laplacian, while acting on functions spanned by these divergenceless harmonic functions. These simultaneous eigen-functions play a central role in determining the propagators of massive Abelian Chern-Simons theory, as well as Maxwell-Chern-Simons theory that we report as a new result, in this paper.

For most part of this paper, we focus on spin-1 fields with boundary conformal dimensions $\Delta = 1$.
Nevertheless, the embedding formalism, with appropriately incorporated parity-odd structures, readily helps
us to write down propagators of parity-violating AdS$_3$ fields with arbitrary spin-$J$ and $\Delta$.
We meticulously record these general results in our paper. However, connecting these results to specific parity violating higher-spin theories in the bulk, will be explored in a future work.

The paper is organised as follows. In section \ref{sec:harm} we discuss the harmonic functions of AdS$_3$ necessary for writing spectral representation of the propagators in AdS. In this section, we also introduce the Chern-Simons operator highlighting its role in the study of parity-odd harmonic functions.
Subsequently, in section \ref{sec:Apps}, we apply the results of the previous sections to study the propagators in various simple theories with a Chern-Simons term.
In both these sections, we have presented the results explicitly in the AdS$_3$ language without any direct use of notations borrowed from the embedding formalism. This
is to ensure an easy comprehension of our work without any familiarity with this formalism. In section \ref{sec:embed}, we review and generalise the embedding formalism to include parity-odd structures and explain the relationship of the results in \eqref{sec:harm} with this formalism. Finally, in section \ref{sec:split}, we present a split representation of the parity harmonic functions, and verify them explicitly in appendix \ref{app:splitverify}.

\noindent
{\bf{Note added:}} While  this manuscript was in preparation \cite{Bhat:2025iqb} appeared on the arXiv which has a small overlap with section \ref{ssec:massCS} of our paper.
%

\section{Harmonic functions in AdS$_3$}\label{sec:harm}
%
In this paper, we will study two-point correlation functions in parity violating QFTs on AdS$_3$. We will work in the Poincar\'e patch of Euclidean AdS$_3$ with the metric
\begin{equation}\label{met}
 ds^2 = \frac{1}{z^2} \left( dz^2 + {dx_1}^2 + {dx_2}^2 \right).
\end{equation}
All the covariant derivatives $\nabla_\mu$ in our analysis will be defined with respect to this metric.
Our main objective is to compute the following bulk-to-bulk propagators of spin-J fields with boundary dimension $\Delta$
\begin{equation}\label{gen2pt}
 {G^{(J,\Delta)}}_{ \alpha_1 \dots \alpha_J \beta_1 \dots \beta_J} (x,y) = \langle h_{\alpha_1 \dots \alpha_J } (x) \, h_{\beta_1 \dots \beta_J}(y)\rangle
\end{equation}
Such propagators in AdS space have been extensively studied in the literature
\cite{Allen:1985wd, Witten:1998qj, DHoker:1999bve, Naqvi:1999va, Leonhardt:2003qu, Leonhardt:2003sn}.
For most part of this paper, we will primarily focus on the following propagator of spin-1 fields $A_\alpha$
\begin{equation}\label{propdef}
 \mathcal G_{\alpha \beta}(x,y) = \langle A_\alpha(x) A_\beta(y) \rangle.
\end{equation}
In particular, we shall compute the bulk correlation functions of gauge fields in abelian Chern-Simons theory,
in a covariant gauge. In this case, the conformal dimensions of the gauge fields will be $\Delta =1$. We will
also consider mass deformations as well as Maxwell-Chern-Simons theory where $\Delta$ is determined by the bulk parameters.
We would like to obtain a spectral representation of these correlation functions \eqref{propdef} and express them in terms of AdS harmonic functions -- eigen functions
of the Laplacian operator in AdS$_3$.
Such a harmonic expansion of \eqref{gen2pt} would allow us to develop the flat space analogue of `momentum-space' Feynman rules in AdS.
The exact form of the parity-even harmonic functions
for arbitrary spin $J$ fields on AdS$_{d+1}$ was worked out in \cite{Costa:2014kfa} using an embedding formalism for AdS. We observe that the general
formalism of \cite{Costa:2014kfa} missed out the possibility of parity-odd harmonic functions, which is particularly relevant for Chern Simons theories in AdS$_3$.
In this section, we generalize their construction for spin-1 AdS harmonics so as to include
new parity-odd functions. We would like to emphasize that the harmonic functions discussed here are limited to those functions which share the same symmetries of
the relevant two-point correlators, and we do not attempt to enumerate the most general set of eigen-functions of the Laplacian.
For the convenience of our readers unfamiliar with the notations of embedding formalism,
in this section, we would like to report our parity-odd generalizations entirely in the $AdS_3$ language, without any reference to the embedding space.
In the theories of our interest, such as Chern-Simons and Maxwell-Chern-Simons theories, the spin-0 and spin-1 harmonic functions are sufficient to express the corresponding propagators. Therefore, in this section, we restrict our discussion to these two cases only. Later on, when we discuss the `embedding formalism' in section \ref{sec:embed}, we shall generalize to higher spins.
%

\subsection{Spin-0 AdS harmonics}
The spin-0 harmonics are scalar eigen functions of the Laplacian in $AdS_3$
\begin{equation}
 \nabla^2 \, \Omega(x,y; \nu) = -\left( \nu^2 +1 \right) \Omega(x,y;\nu) \, ,
\end{equation}
where $\nu$ parametrizes the eigen values.
Here $x=\{z_1,x_1,x_2\}$ and $y=\{z_2,y_1,y_2\}$ are two points in AdS$_3$ parametrized by coordinates in the Poincar\'e patch.
Here $z_{1,2}$ are the corresponding radial coordinates with $z_{1,2} \rightarrow 0$ being the boundary (see the metric in \eqref{met}).
Let us define the chordal distance
\footnote{The geometric meaning of chordal distance $u$ becomes manifest in the embedding space; see section \ref{sec:embed}.} between two points $x$ and $y$ by
\begin{equation}\label{chdis}
 u (x,y) = \frac{(z_1 - z_2)^2 + (x_1 - y_1)^2 + (x_2-y_2)^2}{2 z_1 z_2}.
\end{equation}
While expressing covariant propagators, we need harmonic functions which are a function of $u$ only.
The scalar harmonics functions $\Omega(x,y;\nu) = \Omega(u;\nu)$ are expressed in terms of associated Legendre-$Q$ functions in the following way
\cite{Costa:2014kfa, Carmi:2018qzm}
\begin{equation}\label{scalarhar}
 \Omega(u;\nu) =  \frac{\left( 1 + i \right) \nu  }{4 \pi ^{\frac{5}{2}}
 \left(u (u+2)\right)^{\frac{1}{4}}} ~Q\bigg[1/2,- 1/2 + i \nu  ;u+1 \bigg] 
\end{equation}
These functions satisfy the following completeness relation
\begin{equation}
 \int_{- \infty}^{\infty} d \nu ~\Omega(u;\nu) = \delta \left(x,y\right),
\end{equation}
along with the Orthogonality relation
\begin{equation}
 \int_{\text{AdS}} d^3z ~\Omega(x,z;\nu) \Omega(z,y;\bar{\nu}) = \frac{1}{2}\left(  \delta \left(\nu + \bar{\nu}\right) + \delta \left(\nu - \bar{\nu}\right) \right) \Omega(x,y;\nu),
\end{equation}

\subsection{Spin-1 AdS harmonics: Transverse}
We would like to write down explicit expressions for transverse spin-1 AdS$_3$ harmonics ($\Omega_{\alpha \beta}(x,y; \nu)$) which satisfy
the following condition
\begin{equation}\label{harmcond}
 \begin{split}
  & \nabla^2 \Omega_{\alpha \beta} (x,y; \nu) = -\left( \nu^2 + 2 \right) \Omega_{\alpha \beta} (x,y; \nu), \\
  & \nabla^\alpha \Omega_{\alpha \beta} (x,y; \nu) = 0,
 \end{split}
\end{equation}
where $\nabla$ denotes the covariant derivatives in $x$-coordinates with respect to the metric \eqref{met} acting on vector fields in AdS$_3$.
The first equation implies that $\Omega_{\alpha \beta}(x,y; \nu)$ are the eigen functions of the laplacian operator with $\nu$ parametrizing the eigen values, while the second equation implies that it is transverse i.e. divergence-free.
The eigen value $\nu$ is referred to as the spectral parameter and it is the analogue of momentum for harmonic functions in flat space.

\subsubsection{Parity-even}\label{ssec:poddh}

\noindent
The parity-even harmonics for spin-1 fields in $AdS_3$, which satisfy \eqref{harmcond}, can be expressed as follows
\cite{Costa:2014kfa, Ankur:2023lum}
\begin{equation}\label{evenharm}
 \Omega^{(e)}_{\alpha \beta} (x,y; \nu) = - \frac{\partial^2 u}{\partial x^\alpha \partial y^\beta} ~\Omega_1(u; \nu)
 + \frac{\partial u}{\partial x^\alpha} \frac{\partial u}{\partial y^\beta} ~\Omega_2 (u; \nu).
\end{equation}
where $u$ is the chordal distance \eqref{chdis}.
The functions $\Omega_{1,2}$ in \eqref{evenharm} are expressed in terms of hypergeometric functions as
follows
{\footnote{The normalization of these functions have been fixed by the orthogonality relations \eqref{EEorth} to be discussed later in \ref{ssec:orth}. This is also the exact same normalization which is implied by the representation \eqref{evOpi}. Note that the normalization of $\Pi^{(e)}_{\alpha \beta}$ in \eqref{evOpi} is in turn fixed by the equation of motion \eqref{eompi}. See \cite{Costa:2014kfa} for more details.}}
\begin{equation}\label{O1O2def}
\begin{split}
 \Omega_1(u;\nu) =& \frac{\left(\nu ^2+1\right)}{12 \pi ^2} \left[ 3 ~\, _2F_1\left(1-i \nu ,i \nu +1;\frac{3}{2};-\frac{u}{2}\right)-(u+1) \, _2F_1\left(2-i \nu
   ,i \nu +2;\frac{5}{2};-\frac{u}{2}\right)\right]\\
 \Omega_2(u;\nu) =& \frac{\left(\nu ^2+1\right)}{12 \pi ^2 u (u+2)} \left[ 3 (u+1) ~\, _2F_1\left(1-i \nu ,i \nu +1;\frac{3}{2};-\frac{u}{2}\right) \right. \\ &
\qquad \qquad\qquad\qquad\qquad\qquad
 \left. -(u (u+2)+3) ~\,
   _2F_1\left(2-i \nu ,i \nu +2;\frac{5}{2};-\frac{u}{2}\right)\right]\\
\end{split}
\end{equation}
Following \cite{Ankur:2023lum}, these coefficient functions have been expressed in a form which is
manifestly regular as $u \rightarrow 0$. Also, in this form it is clear that these functions do not have any poles with respect to the spectral parameter $\nu$.
The function $\Omega^{(e)}_{\alpha \beta} (x,y; \nu)$ is manifestly symmetric under simultaneous exchange of the arguments $x$ and $y$ together with an interchange of
the indices
\begin{equation}
 \Omega^{(e)}_{\alpha \beta} (x,y; \nu) = \Omega^{(e)}_{\beta \alpha} (y,x; \nu)
\end{equation}
This property is shared with the correlation functions \eqref{propdef}.

Finally, it is very interesting to note that these coefficient functions can be expressed in terms of the scalar harmonic functions \eqref{scalarhar} in the following way
\begin{equation}\label{Omegrel}
\begin{split}
  \Omega_1(u;\nu) &= - \frac{1}{\nu^2} \left[ u \left( 2+u \right) ~\partial^2_u \, \Omega(u;\nu) + 2\left(1+u\right) ~\partial_u \, \Omega(u;\nu) \right]
  \equiv \mathbb D^{(1)} \left( \frac{1}{\nu^2} \Omega(u;\nu) \right), \\
  \Omega_2(u;\nu) &= - \frac{1}{\nu^2} \left[ \left( 1+u \right) ~\partial^2_u \, \Omega(u;\nu) + 2  ~\partial_u \, \Omega(u;\nu) \right] \equiv \mathbb D^{(2)} \left( \frac{1}{\nu^2} \Omega(u;\nu) \right), \\
     \Omega_1(u, \nu) &= \frac{1}{\nu^2} \left[ (1+\nu^2)\Omega(u, \nu)+(1+u) \ \partial_u \Omega(u, \nu) \right], \\
\end{split}
\end{equation}
where, we have defined the operators $\mathbb D^{1,2}$ involving derivatives of $u$ for future convenience.
These identities turn out to be very useful in the algebraic manipulations involving these harmonic functions.

\subsubsection{Parity-odd}\label{ssec:poddh}
%
We discover that, in addition to $\Omega^{(e)}_{\alpha \beta}$, there exists a set of parity-odd spin-1 harmonic functions which satisfy the conditions \eqref{harmcond}. These functions are given by
{\footnote{Here $\epsilon_{\alpha \rho \sigma}$ is the Levi-civita tensor, where the Levi-civita symbol has been multiplied with the appropriate $\sqrt{g}$ to make it covariant.
}}
\begin{equation}\label{oddharm}
  {\Omega^{(o)}}_{\alpha \beta} (x,y; \nu) = \epsilon_{\alpha}^{~~\sigma \rho} \, \frac{\partial{u}}{\partial x^\rho} \, \frac{\partial^2{u}}{\partial x^\sigma \partial y^\beta} \, \Omega_3 (u ;\nu).
\end{equation}
where
{\footnote{The normalization of the parity-odd harmonics are inferred from that of the even harmonics through the relation \eqref{cseo}.}}
\begin{equation}\label{O3def}
\begin{split}
 \Omega_3 (u ;\nu) &= \partial_u \Omega_1 + \Omega_2 \\
 &= \frac{\nu  \left(\nu  \sqrt{u (u+2)} \cos \left(2 \nu  \sinh ^{-1}\left(\sqrt{\frac{u}{2}}\right)\right)-(u+1) \sin
   \left(2 \nu  \sinh ^{-1}\left(\sqrt{\frac{u}{2}}\right)\right)\right)}{4 \pi ^2 (u (u+2))^{3/2}}.
\end{split}
\end{equation}
It turns out that this expression for $\Omega_3$ is simply a $u$ derivative of the scalar harmonics
\begin{equation}\label{O3Orel}
\Omega_3(u;\nu) = \partial_u \, \Omega(u; \nu).
\end{equation}
Note that, despite appearance, just like its parity even counterpart, these odd harmonics are symmetric under simultaneous exchange of the indices and
the arguments $x$ and $y$
\begin{equation}\label{oexcng}
 {\Omega^{(o)}}_{\beta \alpha} (y,x; \nu) = {\Omega^{(o)}}_{\alpha \beta} (x,y; \nu).
\end{equation}
The Chern-Simons propagator is expected to satisfy a similar property on physical grounds. Therefore, this property of the odd harmonics will enable us to write down a
spectral representation of the Chern-Simons propagator (see section \ref{sec:CSprop}).

\subsubsection{Chern-Simons Operator: Definition and Properties}\label{ssec:CSdef}

We will now introduce a Chern-Simons operator which will play a central role in obtaining most of our results in this paper.
Let $\mathrm d$ denote the exterior derivative, and let $*$ denote the Hodge dual. Then $(* \mathrm d)$ is our Chern-Simons operator.
Now the Hodge Laplacian acting on an one-form $A$ is defined by
\begin{equation}\label{Hodlap}
 \Delta_{\text{Hodge}} A  \equiv *{\mathrm d}*{\mathrm d}A+{\mathrm d}*{\mathrm d}*A ~
 \Rightarrow ~
 \Delta_{\text{Hodge}} A_\mu = - \nabla^2 A_\mu + R_{\mu}{}^{\beta} A_\beta
\end{equation}
If the one form is divergence free, then $(*d*)A = 0$ and hence the second term in the definition
of Hodge Laplacian vanishes. Consequently, for divergence free one-forms,
the Hodge Laplacian reduces to
$\Delta_{\text{Hodge}} A_\mu = (*d)^2A_\mu = - \nabla^2 A_\mu + R_{\mu}{}^{\beta} A_\beta$.
Thus, the Hodge Laplacian is the square of a Chern-Simons operator for divergence-less one-forms. This is a general result,
and  is true for divergence-free one-forms on any manifold of any dimension.

Further, specializing to maximally-symmetric space, we have $R_{\mu \beta} =  2 \Lambda g_{\mu\beta}$, which implies that we can simultaneously diagonalize $\Delta_{\rm Hodge}$ and $\nabla^2$. Also, for divergence-free one-forms on
maximally-symmetric space, we can write
\begin{equation}\label{CSsqr}
 \Delta_{\text{Hodge}} A_\mu = (*d)^2 \, A_\mu = - (\nabla^2 - 2 \Lambda) A_\mu
\end{equation}
In this sense, acting on the space of divergence-free one forms on maximally-symmetric spaces, the Chern-Simons operator squares to the Laplacian.

If we specialize to 3-dimensions, then $A$ and $(*{\mathrm d})A$ are both 1-forms. Hence, in $3$-dimensions we can  act on parity-even harmonic functions with $(*{\mathrm d})$ to obtain parity-odd harmonic functions -- alternatively, we can simultaneously diagonalize $\Delta_{\rm Hodge}$ and $(*{\mathrm d})$, using linear combinations of parity-even and parity-odd harmonics.

We will now record these facts explicitly in the index notation, particularly focusing on the action
of the Chern-Simons operator on the harmonic functions defined in the previous sections. In the index notation, we define
the Chern-Simons operator as
\begin{equation}\label{CSop}
  \mathscr{D}^{\alpha \beta}  \left( \dots \right) \equiv \epsilon^{\alpha \sigma \beta} \nabla_\sigma \left( \dots \right).
\end{equation}
where, again $\nabla$ denotes a covariant derivative. Now, recall that AdS$_3$  with metric
\eqref{met} is a maximally symmetric space with $R_{\mu \beta} = - 2 g_{\mu \beta}$. Note that we have used $\Lambda = -1$ in the Einstein equation which is solved by \eqref{met}.
On the space of divergence free functions $\left( \Omega_{\alpha \beta} \right)$ defined on AdS$_3$, spanned by the parity even and odd harmonics in \eqref{evenharm} and \eqref{oddharm}, the Chern Simons operator squares to the laplacian operator in the following way
\footnote{See \cite{Datta:2011za}, where this property was also identified previously.}
\begin{equation}\label{CSsq}
 (\nabla^2 + 2) \Omega_{\alpha \beta} = - \mathscr D_{\alpha}^{~\mu} \mathscr D_\mu^{~\rho} \Omega_{\rho \beta}.
\end{equation}
In addition, we observe that this Chern Simons operator, converts the parity-even harmonics of \eqref{evenharm} into the parity-odd harmonics of \eqref{oddharm} and vice-versa
\begin{equation}\label{cseo}
\begin{split}
  & \mathscr{D}_{\alpha}^{~\sigma}  ~\Omega^{(e)}_{\sigma \beta} (x,y; \nu) = \Omega^{(o)}_{\alpha \beta} (x,y; \nu) \\
  & \mathscr{D}_{\alpha}^{~\sigma}  ~\Omega^{(o)}_{\sigma \beta} (x,y; \nu) = \nu^2 ~\Omega^{(e)}_{\alpha \beta} (x,y; \nu)
\end{split}
\end{equation}
This allows us to write down a particular linear combination of parity-even and odd spin-1 harmonics which are simultaneously eigen functions of the Chern Simons operator and the Laplacian. This linear combination is given by
\begin{equation}\label{CSeigen}
 \Xi_{\alpha \beta} (x,y; \nu) =  \Omega^{(e)}_{\sigma \beta} + \frac{1}{\nu} \Omega^{(o)}_{\sigma \beta} \, .
\end{equation}
Clearly, $\Xi_{\alpha \beta} (x,y; \nu)$ satisfies \eqref{harmcond}. Also, from \eqref{cseo}, it follows that $\Xi_{\alpha \beta} (x,y; \nu)$ is an eigen function of the Chern Simons operator
with eigen value $\nu$, i.e.
\begin{equation}\label{CSeigeneq}
 \mathscr{D}_{\alpha}^{~\sigma}  ~\Xi_{\sigma \beta} (x,y; \nu) = \nu ~ \Xi_{\alpha \beta} (x,y; \nu).
\end{equation}
Note that the Chern Simons operator squares to the Laplacian as shown in \eqref{CSsq} for a specific class of functions on AdS$_3$, which makes it possible to write down a common set of eigen functions for both the operators. However, all the eigen functions of the Laplacian, which are given by any arbitrary linear combination of even and odd harmonics, are not necessarily an eigen function of the Chern Simons operator. Rather, it is the harmonic functions spanned by $\Xi_{\sigma \beta}$ -- a particular linear combination of odd and even harmonics \eqref{CSeigen}, which are simultaneous eigen functions of the Chern-Simons operator as well as the Laplacian.

Also note that, since both the odd and even harmonics satisfy the exchange property \eqref{oexcng}, we have
\begin{equation}\label{xiexcng}
 {\Xi}_{\beta \alpha} (y,x; \nu) = {\Xi}_{\alpha \beta} (x,y; \nu).
\end{equation}
Before concluding this discussion, let us observe that, our results here can be readily generalized to higher-spin harmonics of the Laplacian on any 3-dimensional maximally-symmetric space. We define the Chern-Simons operator acting on a spin-$J$ field - i.e. a traceless symmetric tensor - as follows,
\begin{equation}
  (\mathscr D \Phi)_{\mu_1 \ldots \mu_J} = \frac{1}{J} \sum_{k=1}^J \epsilon_{\mu_k}{}^{\alpha \beta} \nabla_\alpha \Phi_{\mu_1 \ldots \mu_{k-1} \beta \mu_{k+1} \ldots \mu_J}.
\end{equation}
This reduces to the operator $(* \mathrm d)$ when acting on spin-1 functions, i.e., one-forms.
Squaring this, we get
\footnote{This operator is the Lichnerowicz Laplacian for AdS$_3$ - the analogue of the hodge laplacian acting on traceless symmetric tensors.}
\begin{equation}
(\mathscr D ^2\Phi)_{\mu_1 \ldots \mu_J} = (-\nabla^2 + J(J+1)\Lambda) \Phi_{\mu_1 \ldots \mu_J}.
\end{equation}
Acting with $\mathscr D$ on parity-even spin-$J$ harmonics gives parity-odd spin-$J$ harmonics; and a linear combination of these will simultaneously diagonalize $\mathscr D$ and $\nabla^2$. Explicit expressions for the parity-odd higher-spin harmonics are most conveniently obtained in embedding space, where we can represent $\mathscr D$ through the operator
defined in \eqref{embedCSop} (see section \ref{sec:embed} for further discussions on this operator).

%
\subsection{Spin-1 AdS harmonics: Longitudinal}

The longitudinal spin-1 harmonics, whose divergence does not vanish, are given by the functions
\begin{equation} \label{longmode}
 \Sigma_{\alpha \beta} (x,y;\nu) = \nabla^{(x)}_\alpha \nabla^{(y)}_\beta \Omega \left( u(x,y); \nu \right) =  \frac{\partial^2 u}{\partial x^\alpha \partial y^\beta} ~\partial_u \Omega(u; \nu)
 + \frac{\partial u}{\partial x^\alpha} \frac{\partial u}{\partial y^\beta} ~ \partial^2_u \Omega (u; \nu),
\end{equation}
where $\Omega(u; \nu)$ are the spin-0 harmonics \eqref{scalarhar}. Note that in the first expression, one covariant derivative
is to be taken with respect to the $x$-coordinate while the other is with respect to the $y$-coordinate. Consequently,
both the covariant derivatives are essentially ordinary derivatives since they act on scalar functions. Hence, the second
expression follows straightforwardly.

\noindent
These functions simultaneously satisfy
the following two conditions
\begin{equation}\label{laponlong}
 \begin{split}
  & \left( \nabla^2 + 2  \right) \Sigma_{\alpha \beta} (x,y;\nu) = -\left( \nu^2 + 1 \right) \Sigma_{\alpha \beta} (x,y;\nu) \, , \\
  & \nabla_\alpha \nabla^\mu  \Sigma_{\mu \beta} (x,y;\nu) = -\left( \nu^2 + 1 \right) \Sigma_{\alpha \beta} (x,y;\nu).
  \end{split}
\end{equation}
Hence, $\left( \nabla^2 + 2  \right) \Sigma_{\alpha \beta} (x,y;\nu) = \nabla_\alpha \nabla^\mu  \Sigma_{\mu \beta} (x,y;\nu)$ is
an operator identity on the space of functions spanned by \eqref{longmode}.
It may be noted that these longitudinal harmonics lie in the kernel of the Chern-Simons opeator \eqref{CSop}
\begin{equation}\label{CSonlong}
 \mathscr{D}^{\mu \beta} \Sigma_{\beta \rho} (x,y;\nu) = 0.
\end{equation}
This equation perhaps implies that there are no parity-odd longitudinal harmonic functions.

\subsection{Completeness and Orthogonality for spin-1 harmonics} \label{ssec:orth}

The even spin-1 harmonics satisfy a completeness relation given by \cite{Costa:2014kfa}
\begin{equation}\label{ecomp}
 \int_{-\infty}^{\infty} d \nu  ~ \Omega^{(e)}_{\mu \beta} \left(x,y;\nu \right) + \int_{-\infty}^{\infty} d\nu ~ \frac{\Sigma_{\mu \beta}\left( x,y;\nu \right) }{\nu^2 + 1}
 = g_{\mu \beta} ~\delta \left( x, y \right).
\end{equation}
where $g_{\mu \beta}$ is the AdS$_3$ metric \eqref{met}. Note that both the transverse as well as the longitudinal harmonics are
a part of this completeness relation.
In addition to this completeness relation among even harmonics, we note that the odd harmonics satisfy
\begin{equation}\label{oddzero}
 \int_{-\infty}^{\infty}  d \nu  ~ \frac{1}{\nu} ~\Omega^{(o)}_{\mu \beta} \left(x,y;\nu \right)  = 0 \, .
\end{equation}
This simply follows from the fact that the argument of this integral is an odd function of $\nu$. Putting together \eqref{ecomp} and \eqref{oddzero},
together with the definition \eqref{CSeigen} of eigen function of the Chern-Simons operator, we can conclude
\begin{equation}\label{xicomp}
 \int_{-\infty}^{\infty} d \nu  ~ \Xi_{\mu \beta} \left(x,y;\nu \right) + \int_{-\infty}^{\infty} d\nu ~ \frac{\Sigma_{\mu \beta}\left( x,y;\nu \right) }{\nu^2 + 1}
 = g_{\mu \beta} ~\delta \left( x, y \right).
\end{equation}
This completeness relation will be very useful in writing down propagators in theories with Chern-Simons terms.

Beside the completeness relation, $\Xi_{\mu \beta} \left(x,y;\nu \right)$ also satisfies an orthogonality relation. To see this,
let us start with the orthogonality of the parity-even spin-1 harmonics reported in \cite{Costa:2014kfa}
\begin{equation}\label{EEorth}
\int_{\text{AdS}} \sqrt{g} ~d^3w ~\Omega^{(e)}_{\mu \rho} \left(x,w;\bar{\nu} \right) g^{\rho \sigma} \Omega^{(e)}_{\sigma \beta} \left(w,y;\nu \right)
= \frac{1}{2} \left( \delta \left( \nu + \bar{\nu} \right) + \delta \left( \nu - \bar{\nu} \right) \right) \Omega^{(e)}_{\mu \beta} \left(x,y;\nu \right) \, .
\end{equation}
Acting on this relation with the Chern-Simons operator \eqref{CSop} with respect to the $x$-coordinate, we have
\begin{equation}\label{EOorth}
\int_{\text{AdS}} \sqrt{g} ~d^3w ~\Omega^{(o)}_{\mu \rho} \left(x,w;\bar{\nu} \right) g^{\rho \sigma} \Omega^{(e)}_{\sigma \beta} \left(w,y;\nu \right)
= \frac{1}{2} \left( \delta \left( \nu + \bar{\nu} \right) + \delta \left( \nu - \bar{\nu} \right) \right) \Omega^{(o)}_{\mu \beta} \left(x,y;\nu \right) \, .
\end{equation}
Further, acting with the Chern-Simons operator again, but now with respect to the $y$-coordinate, along with the use of the property
\eqref{oexcng} and \eqref{cseo}, we get
\begin{equation}\label{OOorth}
\int_{\text{AdS}} \sqrt{g} ~d^3w ~\Omega^{(o)}_{\mu \rho} \left(x,w;\bar{\nu} \right) g^{\rho \sigma} \Omega^{(o)}_{\sigma \beta} \left(w,y;\nu \right)
= \frac{\nu^2}{2} \left( \delta \left( \nu + \bar{\nu} \right) + \delta \left( \nu - \bar{\nu} \right) \right) \Omega^{(e)}_{\mu \beta} \left(x,y;\nu \right).
\end{equation}
Finally, putting together \eqref{EEorth}, \eqref{EOorth} and \eqref{OOorth}, together with the definition \eqref{CSeigen}, we obtain
\begin{equation}\label{Xiorth}
\int_{\text{AdS}} \sqrt{g} ~d^3w ~\Xi_{\mu \rho} \left(x,w;\bar{\nu} \right) g^{\rho \sigma} \Xi_{\sigma \beta} \left(w,y;\nu \right)
= \left( \delta \left( \nu + \bar{\nu} \right) + \delta \left( \nu - \bar{\nu} \right) \right) \Xi_{\mu \beta} \left(x,y;\nu \right).
\end{equation}

Equipped with these results we will first proceed to find the Green's function for the Chern Simons operator. Subsequently,
we will further demonstrate the utility of these results by computing the AdS$_3$ propagators for massive Abelian Chern-Simons theory and
Maxwell-Chern-Simons theory.

\subsection{AdS$_3$ harmonics as representations of $SO(3,1)$ isometry}

The generators of isometries of AdS$_3$ must have a well defined action on our harmonic functions (see appendix-\ref{app:gens} for notations and conventions related to these generators). Let us denote the harmonic functions collectively as $\mathcal H (x,y; \nu) = \{ \Omega, \Xi_{\alpha \beta}, \Sigma_{\alpha \beta}\}$. Since our harmonic functions depends on two points, the isometry generators can act as Lie derivatives at both the points. By construction, our harmonic function is invariant under diagonal action of these generators, i.e.
\begin{equation} \label{spin0invar}
 \left( \mathcal L _\xi (x) + \mathcal L _\xi (y) \right) \mathcal H(x,y; \nu) = 0 ,
\end{equation}
for all the killing vectors $\xi$ given in \eqref{so31gen}. However, they transform non-trivially under the action of a single set of generators $\{ \mathcal L _\xi (x) \}$ or $\{ \mathcal L _\xi (y) \}$. The representation with which these functions are constructed can be obtained by the actions of the quadratic Casimirs on these harmonic functions.

We find that, under the action of the Casimir operators at $x$ (or interchangeably at $y$), the spin-0 harmonic functions $\Omega(x,y; \nu)$
transform in the principal series representation with $h = \bar{h} = \frac{1}{2}(1 + i \nu),$ with $\nu \in \mathbb R$. Here $(h,\bar{h})$ is the holomorphic and anti-holomorphic labels used to denote representation of the global 2D conformal algebra.

Similarly, the transverse spin-1 harmonic functions $\Xi_{\alpha \beta}(x,y; \nu)$ also lie in the principal series representation, with
\begin{equation}
 \Xi_{\alpha \beta}(x,y; \nu) :
\begin{cases}
        & h = 1 + i \frac{\nu}{2}, \bar h = i \frac{\nu}{2}, ~~\text{with} ~~\nu > 0. \\
        & h =  i \frac{\nu}{2}, \bar h = 1 + i \frac{\nu}{2}, ~~\text{with} ~~\nu < 0.
\end{cases}
\end{equation}
Also, acting with the Casimir operators on the Longitudinal harmonic functions we get
\begin{equation}
 \Xi_{\alpha \beta}(x,y; \nu) : h = \bar h = \frac{1}{2}\left( 1 + \sqrt{1-\nu^2} \right).
\end{equation}
When $\nu^2 > 1$, this lies in the principal series representation, while for $\nu \in [-1,1]$ this lies in the complementary series.
It is interesting to note that, for these spin-1 harmonics, we find that the Casimir operators $C$ and $\bar C$ defined in \eqref{casimir} are expressed simply
in terms of the Laplacian and the Chern-Simons operator \eqref{CSop}. We find
\begin{equation}
 2 \left( C + \bar C \right) = \nabla^2 + 2 , ~~  C - \bar C  = i \mathscr D.
\end{equation}
Thus, we find that our $\Xi$ defined as a particular combination of the parity-even and parity-odd harmonics in \eqref{CSeigen} diagonalize both the quadratic Casimirs.

The reader must also note that although the harmonic functions are eigen functions of the Casimirs, they are not the eigen function of the Cartan generators $L_0$ and $\bar L_0$, or any other choice of the Cartan-basis. In fact, each harmonic function is constructed out of the entire `principal series multiplet' at $x$ together with those at $y$, such that \eqref{spin0invar} is satisfied\footnote{ This stitching together of `principal series multiplets' is evident in the split representation of our harmonic functions to be discussed in more detail in \ref{sec:split}.}.

\section{Applications of parity-odd spin-1 harmonics in AdS$_3$}\label{sec:Apps}
%
%
%
\subsection{Pure Abelian Chern-Simons propagator in AdS$_3$}\label{sec:CSprop}
%
\subsubsection*{Bulk-to-bulk propagator}
%
The equation for the propagator \eqref{propdef} for the abelian Chern-Simons theory, after fixing a covariant gauge, is given by
\begin{equation}\label{abelEq}
 \left( \mathscr{D}_\mu^{~\alpha} \mathcal  + \frac{1}{\xi} \nabla_{\mu} \nabla^{\alpha} \right) \mathcal G_{\alpha \beta} \left( x,y\right) = -g_{\mu \beta} ~\delta(x,y).
\end{equation}
Here $\xi$ is the gauge fixing parameter of the $R_\xi$ covariant gauge. This equation is easily solved
with the help of the completeness relation \eqref{xicomp}. The solution reads
\begin{equation}\label{abelsol}
 G_{\mu \beta} (x,y) = -\int_{-\infty}^{\infty} d\nu ~ \frac{1}{\nu}~ \Xi_{\mu \beta} \left( x,y; \nu \right)
 + \xi \int_{-\infty}^{\infty} \frac{\Sigma_{\mu \beta} \left(x,y; \nu \right)}{\left( \nu^2 + 1\right)^2}.
\end{equation}
In order to verify that \eqref{abelsol} indeed solves \eqref{abelEq}, we find it convenient to use the relations \eqref{CSeigeneq}, \eqref{CSonlong} \eqref{laponlong}, together with the fact that $\Xi_{\mu \beta}$ is divergence free and satisfies the completeness relation.

\noindent
We find it convenient to work in the Landau gauge where $\xi =0$ and hence the second term drops from the propagator.
Focusing on the first term in \eqref{abelsol} we find

\begin{equation}\label{intsmp}
 \int_{-\infty}^{\infty} d\nu ~ \frac{1}{\nu}~ \Xi_{\mu \beta} =
 \int_{-\infty}^{\infty} d\nu ~ \frac{1}{\nu}~ \left( \Omega^{(e)}_{\mu \beta}
 + \frac{1}{\nu} \Omega^{(o)}_{\mu \beta}  \right) =  \int_{-\infty}^{\infty} d\nu ~ \frac{1}{\nu^2}~ \Omega^{(o)}_{\mu \beta} \, ,
\end{equation}
where the integral over the parity even harmonics vanishes on considering principle value of the $1/\nu$,
since after division by $\nu$, the integrand becomes an odd function of $\nu$.
The left over integral over the parity odd harmonics can be performed explicitly and we get the following
bulk-to-bulk propagator for the abelian Chern-Simons theory
\begin{equation}\label{abelpropres}
  G_{\mu \beta} (x,y) = \frac{1}{4 \pi} ~\epsilon_{\mu}^{~\rho \sigma} \frac{\partial{u}}{\partial x^\rho} \frac{\partial^2{u}}{\partial x^\sigma \partial y^\beta}
  ~ \left( \frac{u+1}{(u (u+2))^{3/2}} \right) \, .
\end{equation}
Since we have an exact expression here, the bulk-to-boundary propagator is found by taking the limit in which one of the bulk point approaches the
boundary. This is a finite limit.

\subsubsection*{Bulk-to-boundary propagator}

The bulk to boundary propagator is obtained by considering the boundary limit of one of the points
of the bulk-to-bulk propagator in \eqref{abelpropres}. More precisely, it is given by
\begin{equation}
 \widetilde G_{\mu \beta} (x, \vec y) = \lim_{\lambda \rightarrow 0}  G_{\mu \beta} (x,\{\lambda z_y,\vec y\}).
\end{equation}
Taking this limit explicitly, we find that the $\widetilde G_{\mu z_y}$ components automatically vanish, and the rest of the
components are given by
\begin{equation}\label{blkbdyprop}
 \widetilde G_{\mu j} (x,\vec{y}) = \frac{1}{4 \pi} ~\epsilon_{\mu}^{~\rho \sigma} \frac{\partial{\tilde u}}{\partial x^\rho} \frac{\partial^2{\tilde u}}{\partial x^\sigma \partial y^j}
  ~ \left( \frac{1}{(\tilde u +1)^2} \right) \, ,
\end{equation}
where $\vec y = \{y_1 , y_2\}$ are the coordinates along the boundary directions and $\tilde u$ is the chordal distance
between a bulk point $x = \{ z_x , x_1 , x_2 \}$ and the boundary point $\vec y$, which is given by
\begin{equation}\label{tildeudef}
\tilde u (x, \vv y)= \frac{1}{2 z_x} \left( \left( x_1 - y_1 \right)^2 + \left( x_2 - y_2 \right)^2 +  {z_x}^2 \right) - 1 \, .
\end{equation}
We can also derive \eqref{blkbdyprop} by projecting the embedding space expression \eqref{embblkbdyodd} to $AdS_3$, after putting $\Delta = 1$.
%
%
We end this discussion by noting that moving to complex coordinates $w = y_1 + i y_2$,  $\widetilde G_{\mu j}$ splits into two components $\widetilde G_{\mu w}$, and  $\widetilde G_{\mu \bar{w}}$. Here 
$\widetilde G_{\mu w}$ matches with the bulk-to-boundary Chern-Simons propagator obtained in \cite{David:2007ak}, when their computation is reduced to the Abelian case.

\subsubsection*{Boundary current two-point function}


The boundary-to-boundary propagator which follows from \eqref{abelpropres} by taking the remaining bulk point to the boundary is given by
\footnote{Note that this boundary-to-boundary propagator can be expressed in terms of Witten's propagator for the Maxwell theory $G_{i j}^{\text{W}} (\vec x, \vec y)$ given in \cite{Witten:1998qj} as follows
\begin{equation}
  G_{i j}^{\text{bdy}} (\vec x, \vec y) = -\frac{1}{2 \pi} \epsilon_{ik} G_{k j}^{\text{W}}.
\end{equation}
}
\begin{equation}\label{bbcs}
 G_{i j}^{\text{bdy}} (\vec x, \vec y) = \frac{1}{2 \pi } \left(   2 \, \frac{\epsilon_{ik} (\vec x- \vec y)_k (\vec x- \vec y)_j}{|\vec x - \vec y|^4} - \frac{\epsilon_{ij}}{|\vec x - \vec y|^2} \right) .
\end{equation}
Shifting to complex variables $z=x_1+i x_2$, and letting $\vec{y} = 0$, \eqref{bbcs} reduces to 
\begin{equation}
  G_{z z}^{\text{bdy}} (z,0) = \frac{i}{ \pi z^2}, ~~ G_{\bar z \bar z}^{\text{bdy}} (\bar z, 0)  =  - \frac{i}{ \pi {\bar{z}}^2},~~
   G_{z \bar z}^{\text{bdy}} = 0  = G_{\bar z z}^{\text{bdy}}.
\end{equation}

Now, following \cite{Kraus:2006nb, David:2007ak, Jensen:2010em}, if we define our Chern-Simons theory with a boundary mass counterterm for the gauge field 
\begin{equation} 
\mathcal S = - \frac{i \kappa}{4 \pi} \int_{\text{AdS}_3} d^3x \,  \epsilon^{\mu \beta \lambda} A_\mu \partial_\beta A_\lambda - \frac{\kappa}{16 \pi} \int_{ \partial \text{AdS} } d^2x  \, A_{i } A^{i}
\end{equation} 
then, we have a holomorphic current on the boundary. The two point function of this holomorphic current, reduces to  (see \cite{David:2007ak} for further details)
$$ \langle J_z J_z\rangle = \frac{\kappa}{ 2  z^2} \, . $$


\subsection{Propagators in Proca theory: A review}\label{ssec:proca}
%
Before proceeding to other theories involving Chern-Simons terms, let us review the parity-even propagator of the massive spin-1 field in the Proca theory.
Our parity-even harmonics can be conveniently expressed in terms of the propagator of this theory. On application of the Chern-Simons operator
on such an identity leads to a new identity for the parity-odd harmonics as well.
The action for the Proca theory is given by
\begin{equation} \label{procaac}
 \mathcal S = \int_{\text{AdS}_3} \sqrt{g} \, dx \, \left( \frac{1}{2} \left( \nabla_\mu A_\beta\right)^2
 - \frac{1}{2}\left( \nabla^\mu A_\mu \right)^2 + M^2 A^\mu A_\mu  \right),
\end{equation}
We shall denote the propagator of this parity-even theory by $\Pi^{(e)}_{\mu \beta}(x,y; \Delta)$ instead of $G_{\mu \beta}$, so that it is easily
highlighted. Also, this notation is similar to that
used in \cite{Costa:2014kfa}, from which we reproduce the result. The theory \eqref{procaac} together with the propagator reduces
to that of the Maxwell theory in the limit $M^2 \rightarrow -2$, with gauge parameter $\xi = 1$.

We will express the mass in \eqref{procaac} in terms of the boundary conformal dimension through the relation $M^2 = \Delta \left( \Delta - 2 \right)-1$. Thus, the propagator $\Pi_{\mu \beta}(x,y; \Delta)$ is a solution to the equation
\begin{equation}\label{eompi}
\left[ \nabla^2 - \left( \Delta \left( \Delta - 2 \right)-1 \right) \right] \Pi^{(e)}_{\mu \beta}(x,y; \Delta)
- \nabla_\mu \nabla^{\alpha} \Pi^{(e)}_{\alpha \beta}(x,y; \Delta) = - g_{\mu \beta} \delta(x,y).
\end{equation}
Solving this equation, the propagator is given by
\begin{equation}\label{fulpropproca}
\Pi^{(e)}_{\alpha \beta}(x,y; \Delta) = \int \, d\nu \, \frac{\Omega_{\alpha \beta}^{(e)}(x,y; \nu)}{\nu^2 + \left( \Delta - 1\right)^2 }
+ \int \, d\nu \, \frac{\Sigma_{\alpha \beta}(x,y; \nu )}{(\nu^2 +1) \left( \Delta -1 \right)^2} \, ,
\end{equation}
where the harmonic functions $\Omega^{(e)}_{\alpha \beta}$ and $\Sigma_{\alpha \beta}$ are defined in \eqref{evenharm} and \eqref{longmode} respectively.
The second integral involving longitudinal harmonics cancel with a part of the answer arising out of the first integral involving transverse harmonics and we obtain
\begin{equation}\label{transproca}
\begin{split}
\Pi^{(e)}_{\alpha \beta}(x,y; \Delta)
= - \frac{\partial^2 u}{\partial x^\alpha \partial y^\beta} ~g_0(u)
 + \frac{\partial u}{\partial x^\alpha} \frac{\partial u}{\partial y^\beta} ~g_1 (u),
\end{split}
\end{equation}
where
\begin{equation}\label{g0g1def}
 \begin{split}
 g_0(u) =& \frac{2^{-\Delta -1} \Delta  u^{-\Delta -1} }{\pi  (\Delta -1)^2} \left((\Delta -2) u ~ _2F_1\left(\Delta -\frac{1}{2},\Delta ;2 \Delta -1;-\frac{2}{u}\right) \right. \\
 & \qquad  \qquad  \qquad \qquad \left. +(u+1) ~ _2F_1\left(\Delta -\frac{1}{2},\Delta +1;2 \Delta -1;-\frac{2}{u}\right)\right) \, , \\
 g_1(u) =& \frac{2^{-\Delta -1} \Delta  u^{-\Delta -2}}{\pi  (\Delta -1)^2 (u+2)}  \left((\Delta -2) u (u+1) \, _2F_1\left(\Delta -\frac{1}{2},\Delta ;2 \Delta -1;-\frac{2}{u}\right)  \right. \\ & \qquad  \qquad  \qquad \qquad\left. +(u (u+2)+3) \, _2F_1\left(\Delta -\frac{1}{2},\Delta +1;2 \Delta -1;-\frac{2}{u}\right)\right)   \, . \\
 \end{split}
 \end{equation}
The corresponding bulk-to-boundary propagator can be defined as
\begin{equation}\label{bdypropdef}
\widetilde \Pi^{(e)}_{\alpha \beta} \left( x,\vec{y}; \Delta \right) = \lim_{z_y \rightarrow 0} \, {z_y}^{1-\Delta} \, \Pi^{(e)}_{\alpha \beta} \left( x,\{ z_y \, , \vec{y} \}; \Delta \right),
\end{equation}
where we have taken the boundary limit on the point $y$ with $y = \{  z_y, \vec{y} \}$, $z_y$ being the radial coordinate while $\vec y$ denotes the boundary coordinates.
Taking this limit on \eqref{transproca}, we get
\begin{equation}\label{bdypropproca}
\widetilde \Pi^{(e)}_{\alpha j} \left( x,\vec{y}; \Delta \right) = \frac{\Delta}{2 \pi (\Delta - 1)}
\left( - \frac{\partial^2 \tilde u}{\partial x^\alpha \partial y^j} ~ \frac{1}{ \left( 2 (\tilde u + 1 ) \right)^\Delta}
 + \frac{\partial \tilde u}{\partial x^\alpha} \frac{\partial \tilde u}{\partial y^j}  ~ \frac{2}{ \left( 2 (\tilde u + 1 ) \right)^{\Delta+1}}\right) \, .
\end{equation}
The contribution of the second integral over longitudinal modes in \eqref{fulpropproca} is subleading with respect to $z_y$ expansion and hence does not contribute to \eqref{bdypropproca}. The form of propagator in \eqref{bdypropproca} is completely consistent with the general form of parity-even bulk-to-boundary propagators predicted by embedding
formalism (see \eqref{genevcorel}). Note that the $\Delta = 1$ corresponds to the Maxwell case, and the presence of the $(\Delta -1)$ factor in the denominator signifies that the Maxwell case
must be treated separately after appropriate gauge fixing (see \cite{Ankur:2023lum} for more details)
\footnote{In $d=3$, for pure Maxwell theory, we can recover the form  \eqref{genevcorel} with $\Delta =1, J = 1$ and $\mathscr C_{\Delta J} = 1/ 2 \pi$, as a coefficient of $\ln (z_y)$.}.

Before concluding this section, we would like to record another related result here.
In the parity-even case, it was shown in \cite{Costa:2014kfa} that the spin-1 harmonics in \eqref{evenharm} can be expressed in terms
of the propagator for the Proca theory \eqref{transproca} in the following way
 \begin{equation}\label{evOpi}
\Omega^{(e)}_{\alpha \beta}(x, y; \nu) =  \frac{i \nu}{2 \pi} \left( \Pi^{(e)}_{\alpha \beta} \left(x, y; 1+i \nu \right)   -  \Pi^{(e)}_{\alpha \beta} \left( x, y ;1 -i \nu \right) \right).
 \end{equation}
This was obtained by performing the integral in the split representation \eqref{ecomp}, which we discuss in more detail in section \ref{sec:split}. Also,
we have verified this using the explicit expressions \eqref{evenharm} and \eqref{transproca}
{\footnote{Note that the $\Omega^{(e)}$ presented in \eqref{O1O2def} has been obtained from \eqref{g0g1def} and \eqref{evOpi}, after using the following hypergeometric identity
{\tiny \begin{equation}
\begin{split}
_2F_1(a,b;c;z) =& \frac{ \Gamma (c) \Gamma (b-a) }{\Gamma (b) \Gamma (c-a)} \, (-z)^{-a} \,  _2F_1\left(a,a-c+1;a-b+1;\frac{1}{z}\right)
+\frac{ \Gamma (c) \Gamma (a-b) }{\Gamma (a) \Gamma (c-b)} \, (-z)^{-b} \,
   _2F_1\left(b,b-c+1;-a+b+1;\frac{1}{z}\right)
\end{split}
\end{equation}
}
}.
We can now act on this relation with the Chern-Simons operator to get
 \begin{equation}\label{oddOpi}
\Omega^{(o)}_{\alpha \beta}(x, y; \nu) =  \frac{i \nu}{2 \pi} \left( \Pi^{(o)}_{\alpha \beta} \left(x, y; 1+i \nu \right)   -  \Pi^{(o)}_{\alpha \beta} \left( x, y ;1 -i \nu \right) \right),
 \end{equation}
where
\begin{equation}
\begin{split}\label{defblkpio}
\Pi^{(o)}_{\alpha \beta} \left( x, y; \Delta \right) &=  \int \, d\nu \, \frac{\Omega_{\alpha \beta}^{(o)}(x,y; \nu)}{\nu^2 + \left( \Delta - 1\right)^2 } \, .
\end{split}
\end{equation}
Unlike its parity-even counterpart, we were unable to find a specific simple parity-odd theory for which \eqref{defblkpio} is the bulk propagator.
The relations \eqref{evOpi} and \eqref{oddOpi} turn out to be useful while performing integrals involving the harmonic functions.

Similar to \eqref{evOpi}, the scalar harmonics \eqref{scalarhar} can also be written in terms of scalar propagators as follows \cite{Carmi:2018qzm}
\begin{equation}\label{scalsplit}
 \Omega(x,y; \nu) = \frac{i \nu}{2 \pi} \left( \Pi(x,y;1+i\nu) - \Pi(x,y; 1-i\nu) \right) \, ,
\end{equation}
where $\Pi$ is the bulk propagator of a massive scalar field with boundary conformal dimension $\Delta$, which is given by
\begin{equation}\label{mascalprop}
 \Pi(x,y; \Delta) = \frac{2^{-\Delta -1} u^{-\Delta } \, _2F_1\left(\Delta ,\frac{1}{2} (2 \Delta -1);2 \Delta -1;-\frac{2}{u}\right)}{\pi } \, .
\end{equation}
It is curious to note that the relation \eqref{evOpi} can be obtained from \eqref{scalsplit} by the use of the identities in \eqref{Omegrel}.

\subsection{Massive Abelian Chern-Simons theory}\label{ssec:massCS}
%
We shall now write down the propagator of a massive Chern-Simons theory given by the action
\begin{equation}
 S = \int_{\text{AdS}_3} \, d^3x \, \left( \frac{i \kappa}{4 \pi} \, \epsilon^{\mu \beta \lambda} A_\mu \partial_\beta A_\lambda  + \frac{1}{2} \, M^2 \,  A^2  \right).
\end{equation}
The propagator of this theory satisfies the equation of motion
\begin{equation}
\frac{i \kappa}{2 \pi} \epsilon_{\alpha}^{~\rho \sigma}  \partial_\rho G_{\sigma \beta} + M^2 G_{\alpha \beta} = - g_{\alpha \beta} \delta(x,y) \, .
\end{equation}
Using a spectral representation, the solution of this equation is given by
\begin{equation}\label{mcsprints}
 G_{\alpha \beta} = - \frac{2 \pi}{i \kappa} \int d \nu \frac{1}{\nu - i \,  m }  \ \Xi_{\alpha \beta} - \frac{1}{M^2} \, \int \, d \nu \, \frac{1}{\nu^2 + 1} \, \Sigma_{\alpha \beta} \, ,
\end{equation}
where $m=2 \pi M^2 / \kappa$.

Let us recall that $\Xi_{\alpha \beta}$ in \eqref{CSeigen} has parity-even and odd pieces
\begin{equation}
 \int d \nu \frac{1}{\nu - i  m} \Xi_{\alpha \beta} =  \int d \nu \frac{1}{\nu - i  m}  \left(\Omega^{(e)}_{\alpha \beta} + \frac{1}{\nu} \Omega^{(o)}_{\alpha \beta} \right).
\end{equation}
We perform the integral of the parity-even and odd parts separately.
In the evaluation of these integrals we have found it useful to use the identities \eqref{Omegrel} and hence perform
the integral over the scalar harmonics. The results can also be verified using the massive spin-1 propagators using \eqref{evOpi}.
For the even part, we get
\begin{equation}\label{mcsevintdef}
  \int d \nu \, \frac{\Omega^{(e)}_{\alpha \beta}}{\nu - i m} =
  - \frac{\partial^2 u}{\partial x^\alpha \partial y^\beta} ~ \mathbb D^{(1)} \left( \mathcal F^{(e)}(u) \right)
 + \frac{\partial u}{\partial x^\alpha} \frac{\partial u}{\partial y^\beta} ~ \mathbb D^{(2)} \left( \mathcal F^{(e)}(u) \right) \, .
\end{equation}
where, the differential operators $\mathbb D^{(1,2)}$ involving derivatives of $u$ are defined in \eqref{Omegrel}. The function $\mathcal F^{(e)}(u)$ obtained after
evaluation of the $\nu$ integral over scalar harmonics is given by
\begin{equation}
 \begin{split}
 \mathcal F^{(e)}(u) = \frac{i ~\text{sgn}(m) \left(1-2^{| m| }
   \left(\sqrt{u}+\sqrt{u+2}\right)^{-2 | m| }\right)}{4 \pi  \sqrt{u
   (u+2)} | m| } \, .
 \end{split}
\end{equation}
Similarly, the integral over the parity-odd harmonic yields
\begin{equation}\label{mcsoa}
\begin{split}
  &\int d \nu \frac{\Omega^{(o)}_{\alpha \beta}}{\nu \left( \nu - i m\right)}
 = \epsilon_{\alpha}^{~~\sigma \rho} \, \frac{\partial{u}}{\partial x^\rho} \, \frac{\partial^2{u}}{\partial x^\sigma \partial y^\beta} \, \mathcal F^{(o)}(u),
   \end{split}
\end{equation}
where
\begin{equation}\label{mcsoafunc}
\begin{split}
\mathcal F^{(o)}(u)
 = -\frac{  2^{|m|} \left(\sqrt{u}+\sqrt{u+2}\right)^{-2 |m|} \left(|m| u (u+2)+\sqrt{u^3 (u+2)}+\sqrt{u (u+2)}\right)}{4 \pi u^2 (u+2)^2} \, .
   \end{split}
\end{equation}
Finally, the second integral over the longitudinal harmonic functions appearing in \eqref{mcsprints} may be performed in a similar fashion, and we obtain
\begin{equation}\label{mcsLintdef}
 \int \, d \nu \, \frac{1}{\nu^2 + 1} \, \Sigma_{\alpha \beta} =
   \frac{\partial^2 u}{\partial x^\alpha \partial y^\beta} ~ \partial_u \left( \mathcal F^{(l)}(u) \right)
 + \frac{\partial u}{\partial x^\alpha} \frac{\partial u}{\partial y^\beta} ~ \partial^2_u \left( \mathcal F^{(l)}(u) \right) \, ,
\end{equation}
where $\mathcal F^{(l)}(u)$ evaluates to
\begin{equation}
\mathcal F^{(l)}(u) = \frac{1}{2 \pi  \sqrt{u (u+2)} \left(\sqrt{u}+\sqrt{u+2}\right)^2}.
\end{equation}
The details related to the computation of these integrals has been incorporated in Appendix \ref{Assec:MCS}.
Putting together \eqref{mcsevintdef}, \eqref{mcsLintdef} and \eqref{mcsoa}, and taking
one of the points to the boundary, we obtain the following bulk-to-boundary propagator
{\footnote{In the $z_y \rightarrow 0$ asymptotic expansion, the term \eqref{mcsLintdef} arising from the integration over the longitudinal harmonics cancels some of the leading terms arising from the transverse piece \eqref{mcsevintdef}.}}
\begin{equation}\label{blkbdymassCS}
\begin{split}
 &\widetilde G_{\alpha j} \left(x , \vec y\right)
 = \lim_{z_y \rightarrow 0} \left[ (z_y)^{-|m|} G_{\alpha \beta}(x,y) \right] \\
 &= \frac{1+|m|}{\kappa \left( 2 (\tilde u +1) \right)^{2 + |m|}} ~ \left[ ~\text{sgn}(m)  \left(   \frac{\partial^2 \tilde u}{\partial x^\alpha \partial y^j}
 \left( 2(\tilde u + 1)\right) - 2 \frac{\partial \tilde u}{\partial x^\alpha} \frac{\partial \tilde u}{\partial y^j}
 \right)
- 2 i \, \epsilon_{\alpha}^{~~\sigma \rho} \frac{\partial{\tilde u}}{\partial x^\rho}
 \frac{\partial^2{\tilde u}}{\partial x^\sigma \partial y^j}
 \right] .
 \end{split}
\end{equation}
Here, the exponent of $(\tilde u +1)$ gives us the conformal dimension of the dual spin-1 operator on the boundary to be $\Delta = 1+ |m|$ (see \eqref{genevcorel} and \eqref{blkbdyJ1delta}).

We can further compute the boundary-to-boundary correlator by moving $x$ to a boundary point using the formula
\begin{equation}
 G^{(b)}_{ij}(\vec x, \vec y) = \lim_{z_x \rightarrow 0} \left[ (z_x)^{-|m|} G_{\alpha j}(x, \vec y) \right] \, .
\end{equation}
Subsequently, after moving to complex coordinates $z = x_1 + i \, x_2$ and setting $\vec y  =0$, we find that only one component of the
two point function is non-zero which, depending on the sign of $m$, is given by
\begin{equation}
 \begin{split}
  & G^{(b)}_{zz} = \frac{1+|m|}{\kappa ~z^{2+|m|} ~ \bar z^{|m|}} , ~~ m < 0 \, ,  \\
  & G^{(b)}_{\bar z \bar z} = \frac{1+|m|}{\kappa ~\bar z^{2+|m|} ~z^{|m|}} , ~~ m > 0 \, .
 \end{split}
\end{equation}
We find that, the massive Chern-Simons theory has single boundary dual operator which is not holomorphic. This operator is a 2D spin-1 operator with scaling dimensions $\Delta = 1 + |m|$. This result matches with that reported in \cite{Bhat:2025iqb}, which corroborates our method. Note that in the $m \rightarrow 0$ limit the answer reduces to either the holomorphic or anti-holomorphic currents of the pure Chern-Simons theory, depending on the sign of $m$.

\subsection{Maxwell-Chern-Simons theories}\label{sec:MCSprop}

Now we consider a Maxwell-Chern-Simons (MCS) theory defined by the action
\begin{equation}\label{maxcsac}
 \mathcal S = \int_{\text{AdS}_3} \, d^3x \, \left( \frac{1}{4 e^2} F_{\mu \beta} F^{\mu \beta}+\frac{i \kappa}{4 \pi} \, \epsilon^{\mu \beta \lambda} A_\mu \partial_\beta A_\lambda + \frac{1}{\xi} \left( \nabla \cdot A \right)^2 \right).
\end{equation}
The last term arises due to gauge fixing, with $\xi$ being the gauge-fixing parameter.
This theory has been studied in \cite{Andrade:2011sx} (also see \cite{Gukov:2004id,Deser:1982vy, Deser:1981wh, Prodanov:1998st}) where the boundary conditions for the classical solutions to this system have been analysed in great detail. These references suggest that MCS theory is a combination of Pure-CS theory and a self-dual massive-CS theory. Here we would like to study MCS theory with the help of the tools developed in this paper.

\noindent
The gauge field propagator in this theory satisfies the following equation of motion
\begin{equation}
 \frac{1}{e^2} (\nabla^2 + 2) G_{\alpha \beta} - \frac{i \kappa}{2 \pi} \epsilon_{\alpha}^{\sigma \rho} \nabla_{\rho} G_{\sigma \beta} - \left( \frac{1}{e^2} - \frac{2}{\xi} \right) \nabla_{\alpha} \nabla^{\mu} G_{\mu \beta} = - g_{\alpha \beta} \delta(x, y).
\end{equation}
The solution of this equation can again be expressed in terms of a spectral representation as follows
\begin{equation}\label{mxcsprop}
G_{\alpha \beta} = e^2 \int d \nu \ \frac{\Xi_{\alpha \beta} }{\nu \left( \nu + i \, \widehat \kappa \right)} + \xi \int d \nu \ \frac{\Sigma_{\alpha \beta} }{2 \left(\nu^2 + 1 \right)^2},
\end{equation}
where $\widehat \kappa = \kappa e^2 / 2 \pi $. The longitudinal piece in this propagator is proportional to $\xi$ indicating
that it is a gauge artifact which can be immediately removed with a choice $\xi = 0$. On the other hand, the poles of the transverse part of the propagator can be split into two parts as follows
\begin{equation}\label{parfracsplit}
 \frac{1}{\nu \left( \nu + i \, \widehat \kappa \right)} = \frac{1}{i \, \widehat \kappa} \left( \frac{1}{\nu} - \frac{1}{\nu +  i \, \widehat \kappa } \right) \, .
\end{equation}
This means that the transverse part of the propagator in \eqref{mxcsprop} is a combination of the same in pure Chern-Simons theory (see \eqref{abelsol}) and that in massive Chern-Simons theory (see \eqref{mcsprints}). Hence the first integral in \eqref{mxcsprop} is a combination
of the answers provided in \eqref{mcsevintdef}, \eqref{mcsoa} and \eqref{abelpropres}.

Let us also recall that the longitudinal part of the propagator was also a gauge artifact for pure Chern-Simons
theory. However, for the massive Chern-Simons theory a longitudinal piece was present in \eqref{mcsprints} - in fact, this theory was not a gauge theory to begin with. Nonetheless, motivated by the split in \eqref{parfracsplit}, let us now investigate whether we can rewrite MCS theory in terms of a pure-CS combined with another theory which resembles a massive-CS theory.

Drawing lessons from \cite{Andrade:2011sx}, let us decompose our gauge field into a flat connection ($a_\mu$) and another field with a non-zero field strength ($b_\mu$), i.e.
\begin{equation}
 A_\mu = a_\mu + b_\mu
\end{equation}
Further, we shall assume that the gauge redundancy of $A_\mu$ entirely migrates to $a_\mu$, which is also a gauge field. This ensures
that $b_\mu$ is a spin-1 field without any gauge redundancy. Also, the classical equation of motion of this theory suggests that $b_\mu$ is a self-dual field (see \cite{Andrade:2011sx}). Substituting this into the action \eqref{maxcsac}, we find that it reduces to
\begin{equation}\label{maxcsac}
 \mathcal S = \int_{\text{AdS}_3} \, d^3x \, \left[ \left( \frac{i \kappa}{4 \pi} ~\epsilon^{\mu \beta \rho} a_\mu \partial_\beta a_\rho + \frac{1}{\xi} \left( \nabla \cdot a \right)^2 \right) +  \left( \frac{1}{4 e^2} F[b]^2 + \frac{i \kappa}{ 4\pi} \, \epsilon^{\mu \beta \rho} b_\mu \partial_\beta b_\rho \right) \right].
\end{equation}
Note that, since $a$ is a flat connection $F[a] =0$, implying $F[a]^2$ vanishes, so does its variation. Although, the CS term itself vanishes for the same reason, its variation does not vanish - it gives rise to a boundary term. For this reason, here we have retained the CS term for $a$ while setting the Maxwell term $F[a]^2$ to zero.  The cross terms between $a$ and $b$ also vanish, up to a boundary term {\footnote{Such a boundary term relating boundary values of $a$ and $b$ would be important if we impose `hybrid' boundary conditions discussed in \cite{Andrade:2011sx}; however, they will not affect our bulk-propagators.}}.

\noindent
Now let us define
\begin{equation}
 G^\mu = \frac{1}{2} \epsilon^{\mu \alpha \beta} F_{\alpha \beta}.
\end{equation}
Then, in terms of $G$ we can rewrite the action as
\begin{equation}\label{maxGcsac}
 \mathcal S \equiv S_a + S_b = \int_{\text{AdS}_3} \, d^3x \, \left[ \left( \frac{i \kappa}{4 \pi} ~\epsilon^{\mu \beta \rho} a_\mu \partial_\beta a_\rho + \frac{1}{\xi} \left( \nabla \cdot a \right)^2 \right) +  \left( \frac{1}{2 e^2} G[b]^2 + \frac{i \kappa}{ 4 \pi} \, \epsilon^{\mu \beta \rho} b_\mu \partial_\beta b_\rho \right) \right].
\end{equation}
Now, we introduce an auxillary field $B$ and rewrite the maxwell term for $b$ as follows
\begin{equation}\label{maxBcsac}
 S_b = \int_{\text{AdS}_3} \, d^3x \, \left(  - \frac{e^2}{2} B_\mu B^\mu - B_\mu G[b]^{\mu}  + \frac{i \kappa}{ 4 \pi} \, ~\epsilon^{\mu \beta \rho} b_\mu \partial_\beta b_\rho  \right) .
\end{equation}
The equation of motion following from this action is
\begin{eqnarray}
 B &=& - \frac{G[b]}{e^2} \label{Beq} \, , \\
 b &=& \frac{2 \pi}{i \kappa} B \label{bbeq} \, .
\end{eqnarray}
If we eliminate $B$ using \eqref{Beq} we get back the Maxwell term in \eqref{maxGcsac}. While if we eliminate $b$ using \eqref{bbeq}
we get
\begin{equation}
 S_b = \int_{\text{AdS}_3} \, d^3x \, \left(  - \frac{e^2}{2} B^2 - \frac{\pi}{i \kappa} ~\epsilon^{\mu \beta \rho} B_\mu \partial_\beta B_\rho   \right) ,
\end{equation}
which resembles the action of a massive CS theory with a pole at $\widehat \kappa$. However, there is one crucial difference. The conditions \eqref{Beq} and \eqref{bbeq}
imply that $B$ is equal to $b$ on-shell and both these fields are self-dual. Which also means that on-shell the $B$ field is transverse.
Using this self-duality condition for the $B$ field we have $\star F [A] = \star F [B] = B$, since $F[a] =0$. Thus, using this condition
we may obtain the correlator for the $B$ field from that of $A$ by acting \eqref{mxcsprop} with suitable Chern-Simons operators. We get
\begin{equation}
\begin{split}
\frac{1}{e^2} \langle B_\alpha B_\beta \rangle  = \frac{1}{e^2} \mathscr D_{\alpha}^{~\mu} \mathscr D_{\beta}^{~\sigma} G _{\mu \sigma} &=   \int d \nu \ \frac{ \nu ~\Xi_{\alpha \beta} }{\left( \nu + i \, \widehat \kappa \right)} \\
& = \int d \nu \ \Xi_{\alpha \beta} - i \widehat \kappa \int d \nu \ \frac{ \Xi_{\alpha \beta} }{\left( \nu + i \, \widehat \kappa \right)} \, .
\end{split}
\end{equation}
Finally, using the identity \eqref{xicomp} for the first integral and dropping the delta-function contact term we get
\begin{equation}
\begin{split}
\frac{1}{e^2} \langle B_\alpha B_\beta \rangle  = - \int d \nu \ \frac{\Sigma_{\alpha \beta}}{(1+ \nu^2)} - i \widehat \kappa \int d \nu \ \frac{ \Xi_{\alpha \beta} }{\left( \nu + i \, \widehat \kappa \right)},
\end{split}
\end{equation}
which is the identical to the propagator for massive CS theory in \eqref{mcsprints}, with the mass identified as $m = - \widehat \kappa$. Hence, this self-dual gauge-invariant part of $A_\mu$ corresponds to a boundary operator with spin-1 and conformal dimension $\Delta = 1 + |\widehat \kappa|$. The flat connection part of $A_\mu$ has a pure CS action (see \eqref{maxGcsac}) and therefore is dual to a chiral current at the boundary as discussed in \ref{sec:CSprop}.

\section{Parity-odd propagators in embedding formalism}\label{sec:embed}
%
The embedding formalism provides us with a powerful framework to simultaneously study AdS propagators and conformal correlators
with significant computational simplification. In the AdS context, this formalism was previously discussed in \cite{Mikhailov:2002bp, Biswas:2002nk, Costa:2014kfa}, which has close parallels with similar techniques introduced in \cite{Costa:2011mg, Costa:2011dw, Penedones:2010ue} for studying conformal correlators. This framework was later developed by several authors and was applied for the study of QFTs in AdS (see for example \cite{Giombi:2017hpr, Carmi:2018qzm, Ankur:2023lum}).
In our paper, many of the results reported above were inspired by this formalism, as we will now discuss.
In fact, we shall translate our results to the embedding formalism language which may be very useful for future applications
of our results. We will begin with a short review of the basics and subsequently, we will explain the consequences of including
parity-odd structures in this formalism.

\subsection{A quick review of basics}
%
Drawing heavily from \cite{Costa:2014kfa}, we will now review some of the basics of the embedding formalism. Although
the formalism has been developed for general dimensions, throughout our discussion here, we will specialize to AdS$_3$
for contextual clarity. Also, we shall present only a bare minimun background to develop an operational familiarity
which will enable our readers to appreciate our novelty of introducing parity-odd structures in this formalism. The reader is
referred to the original references for further details.

In the embedding formalism, Euclidean AdS$_3$ is considered to be hyperbolic hypersurface in four dimensional Minkowski
space $\mathbb{M}^4$. Let us use light-cone coordinates in $\mathbb{M}^4$ with $X = \{ X^+, X^- , \vv{X} \}$, with the metric
\begin{equation}\label{minkmet}
 ds^2_{\mathbb{M}^4} = - \, dX_+ dX_- + \, \delta_{ab} \, dX^a \, dX^b.
\end{equation}
Within $\mathbb{M}^4$, AdS$_3$ is  the hypersurface specified by the constraint
\begin{equation}\label{hyperbola}
  X^2 + 1 = 0 , ~~X_0 \geq 0.
\end{equation}
The Minkowski coordinates are expressed in terms of the AdS$_3$ Poincar\'e coordinates $\{ z , \vv{x} \}$ through the embedding
equations
\begin{equation}\label{embedeq}
 X^+ = \frac{1}{z}, ~~ X^{-} = \frac{z^2 + \vv{x}  \, ^2 }{z}, ~~\vv{X} = \frac{\vv{x} }{z}.
\end{equation}
These embedding equations lead to the induced metric \eqref{met} on the AdS$_3$ hypersurface \eqref{hyperbola}.

The hyperbolic hypersurface \eqref{hyperbola} asyptotically approaches the light-cone
as we tend towards the AdS$_3$ boundary. Therefore, if P denotes
the coordinates of a boundary point, it must be null since it lies on the light-cone. Thus, for a boundary point we can
write
\begin{equation}\label{bdyembed}
 P^2 = 0 : P_+ = 1, ~~P_- = \vv{y}\,^2 , ~~ \vv{P} = \vv{y},
\end{equation}
where $\vv {y}$ are the boundary coordinates of AdS$_3$, and the contraction is performed with the metric \eqref{minkmet}.

\subsubsection*{Embedding Polynomials}

A symmetric traceless AdS$_3$ tensor denoted by
\footnote{Following the conventions in \cite{Costa:2014kfa}, we will denote the AdS$_3$ indices with Greek letters, while
the $\mathbb{M}^4$ indices with capital English aphabets.}
$h_{\alpha_1 \dots \alpha_J}(x)$ can be represented by a polynomial $H(W,X)$ on
$\mathbb{M}^4$. This polynomial is expressed in terms of a tensor on $\mathbb{M}^4$
\begin{equation}\label{polyform}
 H(W,X) = W^{A_1} \dots W^{A_J} H_{A_1 \dots A_J} \left( X\right),
\end{equation}
with $W^2 = 0, X \cdot W = 0$, together with the hypersurface condition \eqref{hyperbola} on $X$.
Note that the lift of an AdS$_3$ tensor to $\mathbb{M}^4$ is not unique. But the behaviour of the
tensor $H_{A_1 \dots A_J} \left( X\right)$ away from the AdS$_3$ submanifold, including
its componenets that are transverse to the hypersurface \eqref{hyperbola}, are not physically relevant for any
conclusions regarding AdS$_3$ tensors. Thus, the tensor is defined to be transverse i.e. $X^{A_1} H_{A_1 \dots A_J} = 0$.
Given this polynomial $H(W,X)$, there is a prescription to recover the corresponding tensor $h_{\alpha_1 \dots \alpha_J}(x)$, which we now review.
At first, we define the following operator which can act on polynomials to generate tensors
 {\footnote{Our convention for $\mathcal K_A$ differs from \cite{Costa:2014kfa} by an overall factor of 2. Also,
 our choice of normalization in \eqref{Kact} is different.}}
 \begin{equation}\label{Kop}
 \begin{split}
  \mathcal K_A  & =  \frac{\partial}{\partial W^A} + X_A \left( X \cdot \frac{\partial}{\partial W} \right) + 2 \left( W \cdot  \frac{\partial}{\partial W} \right) \frac{\partial}{\partial W^A} \\
   &+ 2 X_A \left( W \cdot  \frac{\partial}{\partial W} \right) \left( X \cdot  \frac{\partial}{\partial W} \right)
    - W_A \left( \frac{\partial^2}{\partial W \cdot \partial W} +  \left( X \cdot  \frac{\partial}{\partial W} \right)  \left( X \cdot  \frac{\partial}{\partial W} \right) \right) \, .
 \end{split}
 \end{equation}
%
Then, we can obtain the tensor $H_{A_1 \dots A_J} \left( X\right)$ by acting $\mathcal K _A$ operator on the polynomial $H(W,X)$
as follows
\begin{equation}\label{Kact}
 H_{A_1 \dots A_J} \left( X\right)= \mathcal K_{A_1} \dots \mathcal K _{A_J} ~ H(W,X) \, .
\end{equation}
Finally, the AdS$_3$ tensor $h_{\alpha_1 \dots \alpha_J}(x)$ is obtained by a suitable projection on
 the hypersurface \eqref{hyperbola}
 \begin{equation}\label{blkproj}
  h_{\alpha_1 \dots \alpha_J}(x)  = \frac{\partial X^{A_1} }{\partial x^{\alpha_1}} \dots \frac{\partial X^{A_J} }{\partial x^{\alpha_J}} H_{A_1 \dots A_J} \left( X\right) \bigg|_{X^2 + 1 =0} \, ,
 \end{equation}
 where we should use the embedding equations \eqref{embedeq}.

It is also useful to define another operator in $\mathbb{M}^4$, whose action on the polynomials reduces to action of
a covariant derivative on AdS$_3$ tensors. Such an operator is given by
\begin{equation}\label{covder}
 \mathcal D_A = \frac{\partial}{\partial X^A} + X_A \left( X \cdot \frac{\partial}{\partial X}  \right) + W_A \left( X \cdot \frac{\partial}{\partial W}  \right) \, .
\end{equation}
The divergence of the AdS$_3$ tensor can be obtained from the polynomial by computing $\left(\mathcal D \cdot \mathcal K \right) \, H(W,X)$, while the Laplacian is obtained from  $\left( \mathcal D \cdot \mathcal D \right) H(W,X)$.

Just like the bulk tensors, we can also associate an embedding space polynomial with tensor fields defined on the boundary of AdS$_3$. In this case, for traceless symmetric tensors, the corresponding polynomial is given by
\begin{equation}\label{polyformbdy}
 F(Z,P) = Z^{A_1} \dots Z^{A_J} F_{A_1 \dots A_J} \left( P\right),
\end{equation}
with $P^2 =0,\, Z^2 =0, \, P\cdot Z =0$. In this case, the embedding tensor $F_{A_1 \dots A_J}$  is orthogonal to the light-cone, rather than the hyperbola,
and hence we have $P^{A_1} F_{A_1 \dots A_J} =0$. The boundary analogue of the $\mathcal K_A$ operator
is given by
{\footnote{Note that the expression for $\mathcal K^{(b)}_A$ is given for general dimensions in equation (84) of \cite{Costa:2014kfa}. The
first term in that expression vanishes for $d=2$ i.e. AdS$_3$. However, the vanishing of this coefficient is related to
a choice of normalization for the $\mathcal K^{(b)}_A$ operator. For $d=2$, we must choose a convention where the first term does not vanish.}}
\begin{equation}\label{bdyK}
 \mathcal K^{(b)}_A = \frac{\partial}{\partial Z^A} + \left( Z \cdot \frac{\partial}{\partial Z} \right) \frac{\partial}{\partial Z_A} - \frac{1}{2} \, Z^A \, \frac{\partial^2}{\partial Z \cdot \partial Z} \, .
\end{equation}
Once we have an embedding space tensor, the corresponding boundary tensor is obtained by the projection
 \begin{equation}
  h_{\alpha_1 \dots \alpha_J}(\vv y)  = \frac{\partial P^{A_1} }{\partial y^{\alpha_1}} \dots \frac{\partial P^{A_J} }{\partial y^{\alpha_J}} H_{A_1 \dots A_J} \left( P\right) \bigg|_{P^2 =0}.
 \end{equation}

Henceforth, these prescriptions to recover the bulk and boundary tensors of AdS$_3$ from the embedding polynomials
will be referred to as the `recovery prescription'.

\subsubsection*{Parity-even propagators}

The map between polynomials and tensors discussed above, are particularly useful in understanding the structure of correlation functions, both in AdS and boundary CFT. We are interested in the bulk two-point of functions of spin-$J$ fields with boundary dimension $\Delta$ given in \eqref{gen2pt}).
There is an embedding polynomial which corresponds to ${G^{(J,\Delta)}}_{ \alpha_1 \dots \alpha_J \beta_1 \dots \beta_J} (x,y)$ \, ,
which can be used to constrain and determine this correlation function.
From the embedding space point of view, the basic building blocks of this polynomial are location
coordinates $X$ and $Y$, which corresponds to $x, y$ respectively, through \eqref{embedeq}. Besides, there are
$W_1$ and $W_2$ which denotes the polarization of the two fields
participating in this two-point functions. On the AdS$_3$ hypersurface, these quantities must satisfy
$X^2=-1 = Y^2$, and $X \cdot W_1 = 0 = Y \cdot W_2$.

Now, Lorentz symmetry of the ambient $\mathbb{M}^4$ is $SO(3,1)$, which is also
the isometry of the AdS$_3$ submanifold. In order to ensure general covariance of the
two-point function under this isometry, our polynomial must be Lorentz invariant. In the
\emph{parity-even case}, the only non-trivial SO(3,1) invariant quantities,
which can be constructed out of the four building blocks are $W_1 \cdot W_2$, $X \cdot W_2$, $Y \cdot W_1$ and
$X \cdot Y$. Applying \eqref{embedeq} we see that $X \cdot Y = - (u+1)$, where $u$ is the chordal distance defined in \eqref{chdis}. Therefore, the chordal distance has the geometric interpretation
of being simply related to the geodesic distance in $\mathbb{M}^4$, between the two points in question, which are lying on the
the AdS$_3$ hypersurface. Also, since we wish to obtain two-point function \eqref{gen2pt} from the embedding polynomial
after applying the recovery prescription, the embedding polynomial must be
of degree $J$ in both $W_1$ and $W_2$. All these requirements constrain the structure of such a polynomial
to the following form
\begin{equation}\label{spinJbulkprop}
 \Pi \left( X, W_1; Y, W_2 ; J, \Delta\right) = \sum_{k=0}^{J} \mathcal G_k(u) \left( W_1 \cdot W_2\right)^{J-k}
 \left( \left( X \cdot W_2 \right) \, \left( Y \cdot W_1 \right) \right)^k ,
\end{equation}
where, the functions $\mathcal G_k(u)$ are to be determined by solving the equation of motion satisfied by the propagator.
Hence, the functions $\mathcal G_k(u)$ carry the $\Delta$ dependence.
In this paper, we mostly focus on the spin-1 fields, for which the most general form of the embedding polynomial is
\begin{equation}\label{spin1bulkprop}
 \Pi \left( X_1, W_1; Y, W_2 ; J=1, \Delta \right) = \left( W_1 \cdot W_2\right)  \mathcal G_0(u) + \left( \left( X \cdot W_2 \right) \, \left( Y \cdot W_1 \right) \right) \mathcal G_1(u).
\end{equation}
This propagator, on application of the recovery prescription, reduces to
\begin{equation}\label{spinprop}
 {G^{(J=1,\Delta)}}_{\alpha \beta} (x,y) = - \frac{\partial^2 u}{\partial x^\alpha \partial y^\beta} ~\mathcal G_0 (u)
 + \frac{\partial u}{\partial x^\alpha} \frac{\partial u}{\partial y^\beta} ~\mathcal G_1 (u).
\end{equation}
Again, the coefficient functions are determined form the equation of motion satisfied by this propagator. For example,
if this propagator satisfies \eqref{eompi}, then the coefficient functions must be those given in \eqref{g0g1def}.
The important take away from this discussion is that the most general form of a spin-1 two-point function
is constrained to be of the form \eqref{spinprop}. Further, notice that the
spin-1 harmonic functions, both transverse and longitudinal, reported in \eqref{evenharm} and \eqref{longmode}, also
have the same general structure similar to \eqref{spinprop}.
This helps us to construct the spectral representations
where we can express the propagators in terms of AdS harmonics. With this intention in the hindsight,
the structure of \eqref{spinprop} provides us with a suitable ansatz for seeking AdS harmonics.
As we will see in the next subsection, the structure of spin-1 propagators in \eqref{spinprop} generalizes to include
more terms, once we relax the criterion of parity invariance.

In a similar vein, we can also write down an embedding polynomial for the bulk-to-boundary propagator. The bulk-to-boundary propagator for spin-J field of dimension $\Delta$ can be obtained by taking one of the bulk points in \eqref{spinJbulkprop} to the boundary in the following way
\begin{equation}\label{bdylimit}
 \widetilde \Pi \left( X, W; P, Z ; J, \Delta\right) =
 \lim_{\lambda \rightarrow \infty} \, \lambda^{\Delta} ~ \Pi \left( X, W_1; Y = \lambda P, Z ; J, \Delta \right) \, ,
\end{equation}
where $P$ represents the coordinates of a boundary point and $Z$ is the corresponding boundary polarization.
Now, boundary conformal invariance implies the following scaling relation
\begin{equation}\label{bdyconfcond}
 \widetilde \Pi \left( X, \omega_1 W; \lambda P, \omega_2 Z + \ell P; J, \Delta\right) = \lambda^{- \Delta}
 \left( \omega_1 \omega_2 \right)^J  \widetilde \Pi \left( X, W; P, Z ; J, \Delta\right) \, .
\end{equation}
The only parity-even structure which is consistent with this symmetry is given by
\begin{equation}\label{blkbdyfix}
 \widetilde \Pi \left( X, W; P, Z ; J, \Delta\right) = \mathscr C_{\Delta J}
 \frac{\left[ (-2 P\cdot X) (W \cdot Z) + 2 (W \cdot P )(Z \cdot X)\right]^J}{(-2 P \cdot X )^{\Delta + J}} \, .
\end{equation}
Here the overall constant $\mathscr C_{\Delta J}$ depends on the details of bulk theory satisfied
by the spin-J fields, while the rest of the dependence on the coordinates is entirely fixed by conformal
invariance. This immediately implies the well known result that the corresponding
boundary two-point conformal correlator, which can be obtained from \eqref{blkbdyfix} by taking
the other point to boundary using a limiting procedure similar to \eqref{bdylimit}, is also entirely fixed.
Thus we see that the boundary conformal invariance not only fixes the boundary two-point function, but it also
fixes the bulk-to-boundary propagator up to an overall constant.

It is now straightforward to implement the recovery prescription on \eqref{blkbdyfix} to obtain the
explicit AdS$_3$ form of these bulk-boundary propagators. For arbitrary $J$, the result is slightly cumbersome.
However, for $J=1$, the result can be written down neatly as follows{\footnote{To the best of our knowledge, this explicit form of the bulk-to-boundary propagator has never appeared in the existing literature on this topic.}}
\begin{equation}\label{genevcorel}
\widetilde \Pi \rightarrow \widetilde{ G}_{\alpha j} (x,y; J, \Delta) =
\mathscr C_{\Delta J}
\left( - \frac{\partial^2 \tilde u}{\partial x^\alpha \partial y^j} ~ \frac{1}{ \left( 2 (\tilde u + 1 ) \right)^\Delta}
 + \frac{\partial \tilde u}{\partial x^\alpha} \frac{\partial \tilde u}{\partial y^j}  ~ \frac{2}{ \left( 2 (\tilde u + 1 ) \right)^{\Delta+1}}\right) \, ,
\end{equation}
where $\alpha$ is a bulk index, while $j$ is a boundary index and
$\tilde u$, defined by \eqref{tildeudef}, is the chordal distance between the bulk and boundary points. Note that the parity-even part of the propagator in \eqref{blkbdymassCS} is of this general form.

\subsection{Parity-odd propagators}

If our bulk theory is not parity invariant, then we must include parity-odd structures in the embedding formalism.
For instance, Chern-Simons theories for spin-1 fields constitute one of the most interesting and well studied theories
which violate parity in 3D. In order to study the propagators of such theories we must include parity-odd terms
in the embedding polynomial. The basic constituent of such parity-odd terms in AdS$_3$ is the Levi-Civita tensor.
We observe that the AdS$_3$ Levi-Civita tensor $\epsilon^{\mu \beta \alpha}$ lifts to a very simple $\mathbb M^4$ tensor
\begin{equation}\label{eplift}
 \mathcal E^{BCD} \equiv X_A \, \epsilon^{ABCD} \rightarrow \epsilon^{\mu \beta \alpha},
\end{equation}
where $\epsilon^{ABCD}$ is the $\mathbb M^4$ Levi-Civita tensor.
This tensor is automatically transverse due
to the anti-symmetry of $\epsilon^{ABCD}$.
Upon application of the projection \eqref{blkproj} on $\mathcal E^{BCD}$,
we recover the AdS$_3$ Levi-Civita tensor $\epsilon^{\mu \beta \alpha}$, including the factor of $\sqrt{g}$ which
multiplies the Levi-Civita symbol to convert it into a tensor.

With this new ingredient in our basket, let us revisit the embedding polynomial \eqref{spinJbulkprop} corresponding to the
bulk two-point function of spin-J fields. In addition to the four Lorentz invariant quantities constituted out of $X, W_1$ and $Y, W_2$, which appears in \eqref{spinJbulkprop}, we can also have
\begin{equation}\label{epscal}
 \mathcal E (X,Y,W_1,W_2) \equiv  \epsilon_{ABCD} X^A Y^B W_1^C W_2^D \, .
\end{equation}
Note that this quantity itself is a polynomial in $\mathbb M^4$ and the application of recovery prescription gives us the following familiar AdS$_3$ tensor
\begin{equation}\label{epscalrec}
 \mathcal E (X,Y,W_1,W_2)  \rightarrow  \mathcal E_{A B} = \epsilon_{ABCD}X^C Y^D  \rightarrow  \mathcal E_{\mu \beta}  = \epsilon_{\mu}^{~ \sigma \rho}  \frac{\partial{u}}{\partial x^\rho} \frac{\partial^2{u}}{\partial x^\sigma \partial y^\beta} \, .
\end{equation}
Here in the first step we have obtained a $\mathbb M^4$ tensor by the application of $K^{(1)}_A K^{(2)}_B = \partial_{W_1}\partial_{W_2}$, and in the second step we have applied the projection \eqref{blkproj}.
%

\subsubsection*{Bulk-to-bulk propagators}

Now, upon incorporating this new Lorentz invariant quantity \eqref{epscal}, the embedding polynomial \eqref{spinJbulkprop} generalizes to
\begin{equation}\label{spinJbulkodd}
\begin{split}
 & \Pi \left( X, W_1; Y, W_2 ; J, \Delta\right)  \\
 & = \sum_{n=0}^{J-m} \sum_{m=0}^1 \mathcal G_{(n,m)}(u) \left( W_1 \cdot W_2\right)^{J-(n+m)}
 \left( \left( X \cdot W_2 \right) \, \left( Y \cdot W_1 \right) \right)^{n} \left( \mathcal E (X,Y,W_1,W_2) \right)^m  \, ,
\end{split}
\end{equation}
Note that, in this generalization, the power of $\mathcal E (X,Y,W_1,W_2)$ is either 0 or 1, since it squares to the other two
parity-even factors which has been already considered.
It is straightforward to apply the recovery prescription on this polynomial to recover the
AdS$_3$ form of these bulk propagators.
In particular, for $J=1$, the most general bulk-to-bulk propagator assumes the form
\begin{equation}\label{spin1bulkodd}
\begin{split}
 &\Pi \left( X_1, W_1; Y, W_2 ; J=1, \Delta \right) \\
 &= \mathcal G_{(0,0)}(u) \, \left( W_1 \cdot W_2\right)   +  \mathcal G_{{(1,0)}}(u) \, \left( \left( X \cdot W_2 \right) \, \left( Y \cdot W_1 \right) \right) + \mathcal G_{{(0,1)}}(u) \, \mathcal E (X,Y,W_1,W_2)  .
 \end{split}
\end{equation}
With the recovery prescription this bulk-to-bulk propagator reduces to the explicit form
\begin{equation}\label{spinpropodd}
 {G^{(J=1,\Delta)}}_{\alpha \beta} (x,y) = - \frac{\partial^2 u}{\partial x^\alpha \partial y^\beta} ~\mathcal G_{(0,0)} (u)
 + \frac{\partial u}{\partial x^\alpha} \frac{\partial u}{\partial y^\beta} ~\mathcal G_{(1,0)} (u) + \epsilon_{\alpha}^{~~\sigma \rho} \frac{\partial{u}}{\partial x^\rho} \frac{\partial^2{u}}{\partial x^\sigma \partial y^\beta} ~\mathcal G_{(0,1)} (u).
\end{equation}
The coefficient functions $\mathcal G_{(n,m)}(u)$ depends on the details of the theory. For example,
for pure abelian Chern-Simons theory \eqref{abelEq}, from \eqref{abelpropres} we have
\begin{equation}
 G_{(0,0)} (u) = 0 , ~ G_{(1,0)} (u) = 0, ~ G_{(0,1)} (u) = \frac{1}{4 \pi} \left( \frac{u+1}{(u (u+2))^{3/2}} \right).
\end{equation}
Note the similarity in the structure between the parity-odd term (third term) in the propagator \eqref{spinpropodd} with the parity-odd harmonic functions \eqref{oddharm}.

Before concluding this discussion let us note that it is possible to define the following embedding space operator
which reduces to our Chern-Simons operator \eqref{CSop}
\begin{equation}\label{embedCSop}
 \mathscr D \equiv \left( \epsilon_{ABCD} X^A W_1^B \mathcal D^C \mathcal K_{(1)}^D \right) \left( W_2 \cdot \mathcal K_{(2)}\right) \, ,
\end{equation}
where $\mathcal D_C$ is the covariant derivative \eqref{covder}, which reduces to an ordinary derivative
$\mathcal D_C = \partial_{X^C}$, within this expression. Also, $\mathcal K_{(1,2)}$ are the operators
\eqref{Kop} with respect to $W_1$ and $W_2$ respectively. This Chern-Simons opeator naturally acts on
embedding polynomials. In fact, this operator has the following action on individual terms appearing
in \eqref{spinpropodd}
\begin{equation}\label{CSonspin1}
\begin{split}
   \mathscr D  \left[ g(u) \, \left( W_1 \cdot W_2\right) \right] =& \,
  g'(u) \, \mathcal E (X,Y,W_1,W_2)  \\
   \mathscr D  \left[ g(u) \, \left( \left( X \cdot W_2 \right) \, \left( Y \cdot W_1 \right) \right) \right] =& \,
  g(u) \, \mathcal E (X,Y,W_1,W_2) \\
   \mathscr D  \left[ g(u) \, \mathcal E (X,Y,W_1,W_2) \right] =& \,
    - \left( 2(u+1) g(u) + u(u+2) g'(u) \right) \, \left( W_1 \cdot W_2\right) \\
  & - \left( 2 g(u) + (u+1) g'(u)\right)  \, \left( \left( X \cdot W_2 \right) \, \left( Y \cdot W_1 \right) \right) \, , \\
\end{split}
\end{equation}
where $g(u)$ is an arbitrary function of the chordal distance.
These relations are analogous to the action of the Chern-Simons operator on spin-1 harmonic functions reported in \eqref{cseo}.
%
%
\subsubsection*{Bulk-to-boundary propagators}
%
For parity violating theories, the most general form of the bulk-to-boundary propagator is also modified compared to its parity-even counterpart \eqref{blkbdyfix}.
This most general expression now must also include parity-violating structures.
Taking into account constraints from boundary conformal invariance \eqref{bdyconfcond}, the
most general bulk-to-boundary for spin-J field with boundary conformal dimension $\Delta$ is
given by
\begin{equation}\label{blkbdygen}
 \widetilde \Pi \left( X, W; P, Z ; J, \Delta\right)
 =  \mathscr C^{(e)}_{(\Delta,J)} ~\widetilde \Pi^{(e)} \left( X, W; P, Z ; J, \Delta\right) +
   \mathscr C^{(o)}_{(\Delta,J)}  ~\widetilde \Pi^{(o)} \left( X, W; P, Z ; J, \Delta\right),
\end{equation}
where the even part is given by a term similar to \eqref{blkbdyfix}
\begin{equation}\label{blkbdyeven}
 \widetilde \Pi^{(e)} \left( X, W; P, Z ; J, \Delta\right) =
 \frac{\left[ (-2 P\cdot X) (W \cdot Z) + 2 (W \cdot P )(Z \cdot X)\right]^J}{(-2 P \cdot X )^{\Delta + J}},
\end{equation}
while the odd term is a new structure
\begin{equation}\label{blkbdyodd}
 \widetilde \Pi^{(o)} \left( X, W; P, Z ; J, \Delta\right) = - \frac{\Delta}{\pi}
 \frac{\left[ (-2 P\cdot X) (W \cdot Z) + 2 (W \cdot P )(Z \cdot X)\right]^{\left(J-1\right)} \mathcal E (X,W,P,Z) }{(-2 P \cdot X )^{\Delta + J}}.
\end{equation}
Here, $\mathscr C^{(e,o)}_{(\Delta,J)}$ are non-universal constants which depends on the details of the bulk theory. Also,
we have chosen a particular normalization for $\widetilde \Pi^{(o)}$ so as to be readily compatible with other conventions in this paper.
For $J=1$, $\widetilde \Pi^{(o)}$ takes a very simple form
\begin{equation}\label{embblkbdyodd}
\widetilde \Pi^{(o)} \left( X,P,W,Z; J=1, \Delta \right)  = \left(- \frac{\Delta}{\pi} \right) \frac{\epsilon^{ABCD}X_A P_B W_C Z_D}{\left(-2 P \cdot X \right)^{\Delta + 1}} \, .
\end{equation}
The corresponding $AdS_3$ quantity obtained after applying the recovery prescription is given by
\begin{equation}\label{blkbdyJ1delta}
 \widetilde G^{(o)}_{\mu j} \left(x , \vec y\right) =  \frac{\Delta}{ \pi} ~\epsilon_{\mu}^{~~\rho \sigma} \frac{\partial{\tilde u}}{\partial x^\rho} \frac{\partial^2{\tilde u}}{\partial x^\sigma \partial y^j}
  ~ \left( \frac{1}{ \left( 2 (\tilde u +1) \right)^{\Delta + 1}} \right) \, ,
\end{equation}
where $\vec y = \{y_1 , y_2\}$ are the coordinates along the boundary directions and $\tilde u$ is the chordal distance
between a bulk point $x = \{ z_x , x_1 , x_2 \}$ and the boundary point $\vec y$, which is given by \eqref{tildeudef}.
For $\Delta = 1$ this quantity reduces to the bulk-to-boundary Chern-Simons propagator \eqref{blkbdyprop}. While, for arbitrary
$\Delta$ this quantity appears in \eqref{blkbdyoddDelta} which is helpful for the construction of the split representation of the Chern-Simons propagator.
Despite the simplicity, we were unable to find a simple bulk theory for a spin-1 field with arbitrary $\Delta$, whose bulk-to-boundary propagator reduces to \eqref{blkbdyJ1delta}.

Nevertheless, we can take the bulk point $x$ in \eqref{blkbdyJ1delta}
to the boundary, which  yields the following boundary-to-boundary propagator, after dropping an overall constant
\begin{equation}\label{blkbdyJ1deltaex}
 \widetilde G^{(o)}_{i j} \left(\vv x , \vv y \right) =  \left(  2 \, \frac{\epsilon_{ik} (\vec x - \vec y)_k (\vec x- \vec y)_j}{|\vec x - \vec y|^{2(\Delta+1)}}
 - \frac{\epsilon_{ij}}{|\vec x - \vec y|^{2 \Delta}} \right) \, .
\end{equation}
This is the most general parity-odd contribution to the
two-point function of any 2D operator with spin-$1$ and conformal dimension $\Delta$.

It is also tempting to ponder whether there is a simple parity-odd higher spin bulk theory which gives rise to a bulk-to-boundary propagator of the form \eqref{blkbdyodd}, particularly for $\Delta = 1$. A higher spin Chern-Simons theory may be a possibility (see section \ref{disco} for further discussion on this point). We defer a careful examination of this possibility for future work.

\section{Split representation of Chern-Simons propagators}\label{sec:split}
%
The idea of a split representation goes back to \cite{Leonhardt:2003qu}.
It was demonstrated in \cite{Costa:2014kfa} that the bulk-to-bulk propagators of fields with arbitrary spin can be expressed as product of
bulk-to-boundary propagators, with an integration over the boundary point
(also see \cite{Fitzpatrick:2011ia} for similar results in the context of AdS scattering amplitudes). In particular,
such a split representation is well studied for spin-1 fields including that for the Maxwell theory \cite{Ankur:2023lum, Giombi:2017hpr}.
We find that a similar construction is also possible for our Chern Simons propagator.
It was shown in \cite{Costa:2014kfa} that the parity even harmonic functions $\Omega_{\alpha \beta}^{(e)}(x,y; \nu)$ can be expressed
in terms of the bulk-to-boundary propagator of the Proca theory \eqref{bdypropdef}, through the following relationship
\begin{equation}\label{ecomp}
\Omega^{(e)}_{\alpha \beta}(x, w; \nu) =  \frac{\nu^2}{\pi} \int_{\partial AdS_3} d^2 y  ~ \widetilde \Pi^{(e)}_{\alpha j} \left(x,\vec y; 1+i \nu \right) ~ \delta^{jk} ~\widetilde \Pi^{(e)}_{k \beta} \left(\vec y, w ;1 -i \nu \right).
 \end{equation}
where $j$ and $k$ are indices corresponding to the AdS boundary directions and is summed with the flat metric.

\noindent
By consecutive action of the Chern-Simons operator \eqref{CSop} on the $x$ coordinate and $w$ coordinate respectively
we obtain the following identities
\begin{equation}\label{split1}
\begin{split}
 \Omega^{(o)}_{\lambda \beta}(x,w; \nu) &= \frac{\nu^2}{\pi} \int_{\partial AdS_3} d^2 y  ~\Pi^{(o)}_{\lambda j} \left(x,y; 1+i \nu \right) ~ \delta^{jk} ~ \widetilde \Pi^{(e)}_{k \beta} \left(y,w; 1- i \nu \right), \\
  \ \Omega^{(e)}_{\lambda \gamma}(x,w; \nu) &= \frac{1}{\pi} \int_{\partial AdS_3} d^2 y  ~\Pi^{(o)}_{\lambda j} \left(x,y; 1+i \nu \right) ~ \delta^{jk} ~ \Pi^{(o)}_{k \gamma} \left(y,w; 1- i \nu \right).
\end{split}
 \end{equation}
where we have again used \eqref{cseo} along with the property \eqref{oexcng}. The quantities $\Pi^{(o)}_{\lambda j}$ used in this expression
are defined by
\footnote{We have checked that the Chern Simons operator commutes with the boundary limit.}
\begin{equation}
\begin{split}\label{blkbdyoddDelta}
\widetilde \Pi^{(o)}_{\alpha \beta} \left( x,\vec{y}; \Delta \right) &= \mathscr D_{\alpha}^{~\mu}\left( \lim_{\lambda \rightarrow 0} \, \lambda^{1-\Delta} \, \Pi^{(e)}_{\mu \beta} \left( x,\{ \lambda \, z_y \, , \vec{y} \}; \Delta \right) \right)  \\
&= \lim_{\lambda \rightarrow 0} \, \lambda^{1-\Delta} \, \int \, d\nu \, \frac{\Omega_{\alpha \beta}^{(o)}(x,\{ \lambda \, z_y \, , \vec{y} \}; \nu)}{\nu^2 + \left( \Delta - 1\right)^2 } \, .
\end{split}
\end{equation}
It is important to note that $\widetilde \Pi^{(o)}_{\alpha \beta}$ is not the bulk-to-boundary Chern-Simons propagator \eqref{blkbdyprop}. However, it is a
quantity which is expressed entirely in terms of the parity odd vector harmonic functions.
In fact, we were not able to write down a simple theory for which this was the propagator. Nonetheless, it enjoys
all the properties similar to a propagator. In appendix \ref{app:splitverify}, we provide a direct verification of
\eqref{split1}, by performing the boundary integral, with the help of the embedding formalism.

Putting together these results in the definition \eqref{CSeigen} we obtain
\begin{equation}\label{ecompxi}
\Xi_{\alpha \beta}(x, w; \nu) =  \frac{\nu^2}{\pi} \int_{\partial AdS_3} d^2 y  ~ \widetilde \Theta_{\alpha j} \left(x,\vec y; 1+i \nu \right) ~ \delta^{jk} ~\widetilde \Theta_{k \beta} \left(\vec y, w ;1 -i \nu \right).
\end{equation}
where $\widetilde \Theta$ is given by
\begin{equation}
\widetilde \Theta =   \widetilde \Pi^{(e)}_{\alpha j} + \frac{1}{\nu} \widetilde \Pi^{(o)}_{\alpha j} \, .
\end{equation}

Using this expression for the
harmonic function into the definition of the bulk-to-bulk propagator in Chern-Simons theory \eqref{abelsol} (choosing $\xi =0$ gauge), we get
\begin{equation}
  G_{\mu \beta} (x,w) = \int_{-\infty}^{\infty} \, d\nu ~ \frac{\nu}{\pi} \int_{\partial AdS_3} d^2 y  ~ \widetilde \Theta_{\alpha j} \left(x,\vec y; 1+i \nu \right) ~ \delta^{jk} ~\widetilde \Theta_{k \beta} \left(\vec y, w ;1 -i \nu \right).
\end{equation}
This is the split representation for our bulk Chern-Simons propagator. This representation may be very useful while computing
Witten diagrams with loops \cite{Giombi:2017hpr, Sleight:2017cax}, in interactive theories with a Chern-Simons term.

\section{Discussions}\label{disco}

In this paper, we have developed a generalization of the embedding formalism to include parity odd structures which enhances the scope of this framework to
study parity violating QFTs in AdS$_3$. Our construction of
parity odd harmonic functions is expected to be extremely valuable for a perturbative
analysis of such theories. Further, the spectral representation of propagators in terms of our harmonic functions, together with their split representation is expected
to provide a significant computational advantage in the evaluation of Witten diagram with loop, while performing a perturbative analysis in such QFTs.

One of the highlights of our work is the recognition of the importance of the
Chern-Simons operator and its role played in relating the parity-odd and even structures appearing in our analysis. It has played a crucial role in seamlessly
integrating the parity-odd terms with their parity-even counterparts. For most part of this paper, we have focused on spin-1 fields in AdS$_3$ with boudary dimension $\Delta =1$. As an application of our results, we were able to write down explicit expressions for the propagators in (massive) Chern-Simons theories and Maxwell-Chern-Simons theories
with significant ease, demonstrating the computational advantage provided by this framework. Another advantage of our formalism is that it is completely covariant which enables us to use the full isometry of AdS. This would be particularly helpful in taking flat space limit of the results \cite{Gadde:2022ghy},
which can provide a non-trivial check on the answers while dealing with more complicated interactive theories in AdS.

Another important development in our paper was the generalization of the
Chern-Simons operator for arbitrary spin, which retains all the nice
properties of its spin-1 counterpart. This operator, together with the scope of
the embedding formalism, immediately generalizes all our results in AdS$_3$ to arbitrary spin and $\Delta$. In the future, it would be interesting to apply our framework to study higher-spin parity violating theories. A particularly simple class of such higher-spin theories in AdS$_3$ is the Abelian Chern-Simons theories for
spin-J fields given by the action \cite{Tyutin:1997yn, Datta:2011za}
\begin{equation}
 S =  \frac{i \kappa}{4 \pi} \int_{\text{AdS}_3} \, \sqrt{g} \, d^3x \,~ h_{\alpha_1}^{~~~\alpha_2 \dots \alpha_J} \epsilon^{\alpha_1 \beta \gamma} \nabla_{\beta} h_{\gamma \alpha_2 \dots \alpha_J} \, .
\end{equation}
It is tempting to conclude that the propagator of these higher spin fields with $\Delta = 1$ is given by \eqref{blkbdyodd}. A comprehensive study of these theories is left for future work.

Another immediate application of our results is to analyse non-Abelian
Chern-Simons theories in AdS$_3$. Given the abelian result, the propagator for pure
non-Abelian Chern-Simons theory is straightforward to write down
%
\begin{equation}\label{nonabel}
  \langle A^a_{\mu} A^b_{\beta} \rangle = \frac{d^{ab}}{4 \pi} ~\epsilon_{\mu}^{~ \sigma \rho} \frac{\partial{u}}{\partial x^\rho} \frac{\partial^2{u}}{\partial x^\sigma \partial y^\beta}
  ~ \left( \frac{u+1}{(u (u+2))^{3/2}} \right) \, .
\end{equation}
Such non-Abelian theories have a 3-point interaction vertex as well. In this theory,
it would be interesting to evaluate Witten diagrams and reproduce the boundary
current correlators verifying its consistency with the boundary current algebra. In fact, it would also be very interesting to perturbatively verify the non-renormalization theorems in this theory by demonstrating the cancellation between
gluon and ghost loops. Further, such abelian theories can be coupled with vector matter, which is expected to possess a rich class of non-susy dualities following lessons from flat space. Our results  can be readily applied to obtain loop corrected propagators and correlation functions in such theories, with which we can corroborate whether such dualities continue to hold in AdS$_3$. Our formalism is expected to provide a clear comprehension of the implications of such duality for the corresponding 2D boundary CFTs. It is tempting to expect that a novel bosonization in 2D CFTs would follow from the bulk dualities between bosonic and fermionic theories. It would be very interesting to investigate whether such a bosonization corresponds to the usual 2D bosonization or it is an entirely new mechanism of bosonization. We hope to report on these questions in another publication in the near future.

\acknowledgments
We would like to thank Sourav Singha for initial collaboration and several useful inputs.
We thank Ant\'onio Antunes, Arjun Bagchi, Diptarka Das, Kaushik Ghosh, Sayan Kar, Apratim Kaviraj, Nilay Kundu, Swapnamay Mondal, Shiraz Minwalla, Omkar Nippanikar, S. Pratik Khastgir, Kaushik Ray, Arnab Priya Saha and  Shubham Sinha for several useful discussion and valuable inputs. We are grateful to Justin David for valuable comments on the draft of this manuscript. JB is grateful for hospitality at IIT Kanpur during the workshop `Aspects of CFT - 2', where part of this work was carried out. SP acknowledges the support of DST-SERB grant (CRG/2021/009137). AS would like to acknowledge the support from Prime Minister’s Research Fellowship (\href{https://www.pmrf.in/}{PMRF}), offered by the Government of India.

%
\appendix
%
%

\section{Explicit verification of split representations}\label{app:splitverify}
%
In this appendix, we report the explicit verification of the split representations of parity-even and odd spin-1 harmonic functions
which was reported in section \ref{sec:split}.

The scalar harmonics functions in \eqref{scalarhar} also has a split representation, which enables us to write them as a boundary integral of a product of two bulk-to-boundary propagators of a massive scalar theory \eqref{mascalprop}. In the
embedding  formalism this statement can be summarized as follows \cite{Carmi:2018qzm}
\begin{equation} \label{scalarsplit}
\begin{split}
\Omega(u, \nu) = \frac{\nu^2}{4 \pi^3}  \int_{\partial} d P  \frac{1}{(-2 X_1.P)^{1+i\nu} (-2 X_2.P)^{1-i\nu}} ,
\end{split}
\end{equation}
We shall employ this result in our computations below.

\subsection{Parity-even harmonics: even-even split}
%
The split representation of the parity-even harmonic functions in \eqref{ecomp}, was reported and proved in \cite{Costa:2014kfa}. We shall
revisit that proof here since similar steps can be suitably adapted for the split representations involving parity-odd harmonics as well.
We begin with the embedding space expression for the bulk-boundary propagator for
\begin{equation} \label{spin1bulkbdy}
\widetilde \Pi^{(e)}(X, P; W, Z; \Delta) = C_{\Delta}\frac{ (-2X . P)(W.Z) + 2 (W.P) (Z.X)  }{(-2 X.P)^{\Delta + 1}}, \qquad C_{\Delta} = \frac{\Delta}{2 \pi (\Delta-1)} \, .
\end{equation}
We are interested in evaluating the following integral
\begin{equation} \label{reqdinteven}
\mathscr{I} = \frac{\nu^2}{\pi} \ \int_{\partial} d P \ \widetilde \Pi^{(e)}(X_1, P; W_1, K^{(b)}_Z; 1+ i \nu) \widetilde \Pi^{(e)}(P, X_2; Z, W_2; 1- i \nu) \, ,
\end{equation}
where we can use $K^{(b)}_A = \frac{\partial }{\partial Z^A}$. The integral \eqref{reqdinteven} then is written as
\begin{equation} \label{reqdintee1}
\begin{split}
& \frac{1+\nu^2}{4 \pi^3} \int_{\partial} d P \frac{ (-2X_1 . P)(W_1. \frac{\partial }{\partial Z}) + 2 (W_1.P) (X_1. \frac{\partial}{\partial Z})  }{(-2 X_1.P)^{2+i \nu}}  \ \frac{ (-2X_2 . P)(W_2.Z) + 2 (W_2.P) (Z.X_2)  }{(-2 X_2.P)^{2-i \nu}} \\ & = \frac{1+\nu^2}{4 \pi^3} \int_{\partial} d P  \Bigg( (-2 X_1.P)(-2 X_2.P)(W_1.W_2) +  2 (-2 X_1.P) (W_2.P)(W_1.X_2) + \\ & ~~2 (W_1.P) (-2 X_2.P) (W_2.X_1) + 4 (W_1.P)(W_2.P)(X_1.X_2) \Bigg) \frac{1}{(-2 X_1.P)^{2+i \nu}(-2 X_2.P)^{2-i \nu}} \, .
\end{split}
\end{equation}
The integral of the term
\begin{equation} \label{eefirst}
\begin{split}
\mathscr{I}_1 &=  \int_{\partial} d P \frac{(-2 X_1.P)(-2 X_2.P)(W_1.W_2)}{(-2 X_1.P)^{2+i \nu}(-2 X_2.P)^{2-i \nu}} \\ & = (W_1.W_2) \int_{\partial} d P \frac{1}{(-2 X_1.P)^{1+i \nu}(-2 X_2.P)^{1-i \nu}} = \frac{4 \pi^3}{\nu^2}(W_1.W_2) \Omega(u,\nu) \, ,
\end{split}
\end{equation}
where we have used \eqref{scalarsplit} to evaluate \eqref{eefirst}. The integral
\begin{equation} \label{eesec1}
\begin{split}
\mathscr{I}_2  & =    \int_{\partial} d P \frac{2 (-2 X_1.P)(W_2.P)(W_1.X_2)}{(-2 X_1.P)^{2+i \nu}(-2 X_2.P)^{2-i \nu}} =  2 (W_1.X_2) \int_{\partial} d P \frac{(W_2.P)}{(-2 X_1.P)^{1+i \nu}(-2 X_2.P)^{2-i \nu}}  \\& =  2 (W_1.X_2) W_{2\alpha} \int_{\partial} d P \frac{P^{\alpha}}{(-2 X_1.P)^{1+i \nu}(-2 X_2.P)^{2-i \nu}}  \\& =  \frac{2 (W_1.X_2)}{2 (1-i \nu)} W_{2\alpha}  \frac{\partial }{\partial X_{2\alpha}} \int_{\partial} d P \frac{1}{(-2 X_1.P)^{1+i \nu}(-2 X_2.P)^{1-i \nu}} \, .
\end{split}
\end{equation}
Using \eqref{scalarsplit}, we can write \eqref{eesec1} as
\begin{equation} \label{eesec2}
\begin{split}
\mathscr{I}_2 =  \frac{4 \pi^3 (W_1.X_2)}{\nu^2 (1-i \nu)} W_{2\alpha}  \frac{\partial }{\partial X_{2\alpha}} \Omega(u,\nu) = \frac{4 \pi^3 (W_1.X_2)}{\nu^2 (1-i \nu)} W_{2\alpha}  \frac{\partial u}{\partial X_{2\alpha}} \partial_u \Omega(u,\nu) \, ,
\end{split}
\end{equation}
where $u = -1-X_1.X_2$ is the chordal distance between two points $X_1$ and $X_2$. Using $\frac{\partial u}{ \partial X_{2\alpha}} = - X_1^{\alpha} $, we find
\begin{equation} \label{eesec3}
\begin{split}
\mathscr{I}_2 = -\frac{4 \pi^3 (W_1.X_2)(W_2.X_1)}{\nu^2 (1-i \nu)} \partial_u \Omega(u,\nu) \, .
\end{split}
\end{equation}
The expression of the integral
\begin{equation} \label{eethird1}
\mathscr{I}_3
=   \int_{\partial} d P \frac{2 (-2 X_2.P)(W_1.P)(W_2.X_1)}{(-2 X_1.P)^{2+i \nu}(-2 X_2.P)^{2-i \nu}}
\end{equation}
can be obtained from integral \eqref{eesec1} by changing $1 \leftrightarrow 2 $ along with $\nu \rightarrow -\nu$. The same transformation can be done in \eqref{eesec3} on terms apart from the derivative of the scalar harmonics, to get the result
\begin{equation} \label{eethird2}
\begin{split}
\mathscr{I}_3 = -\frac{4 \pi^3 (W_1.X_2)(W_2.X_1)}{\nu^2 (1+i \nu)} \partial_u \Omega(u,\nu).
\end{split}
\end{equation}
The final term that we need to integrate is
\begin{equation} \label{eefourth1}
\begin{split}
\mathscr{I}_4 & =  4 \int_{\partial} d P \frac{(W_1.P)(W_2.P)(X_1.X_2)}{(-2 X_1.P)^{2+i \nu}(-2 X_2.P)^{2-i \nu}} \\& = 4 (X_1.X_2) W_{1\alpha} W_{2\alpha}  \int_{\partial} d P \frac{P^{\alpha}P^{\beta}}{ (-2 X_1.P)^{2+i \nu}(-2 X_2.P)^{2-i \nu}} \\ & = \frac{4 (X_1.X_2) W_{1\alpha}W_{2\alpha}}{4(1+i\nu)(1-i\nu)} \frac{\partial}{\partial X_{1\alpha}} \frac{\partial}{\partial X_{2\alpha}} \int d P  \frac{1}{(-2 X_1.P)^{2+i \nu}(-2 X_2.P)^{2-i \nu}} \\& = \frac{4 \pi^3 (X_1.X_2)}{\nu^2(1+\nu^2)}W_{1\alpha}W_{2\alpha}  \frac{\partial}{\partial X_{1\alpha}} \frac{\partial}{\partial X_{2\alpha}} \Omega(u, \nu) \\& = \frac{4 \pi^3}{\nu^2(1+\nu^2)} (X_1.X_2)W_{1\alpha}W_{2\alpha}  \left[ \frac{\partial^2 u}{\partial X_{1\alpha} \partial X_{2\beta}} \partial_u \Omega(u, \nu) + \frac{\partial u}{\partial X_{1\alpha}} \frac{\partial u}{ \partial X_{2\beta}} \partial^2_u \Omega(u, \nu)  \right] \, .
\end{split}
\end{equation}
The chordal distance in terms of the points $X_1$ and $X_2$ is given as $u = -1 - X_1 . X_2$. Therefore, we find that
\begin{equation}\label{eefourth2}
\frac{\partial u}{\partial X_{1\alpha}} = - X_2^{\alpha}, \  \frac{\partial u}{\partial X_{2\beta}} = - X_1^{\beta}, \ \frac{\partial^2 u}{\partial X_{1\alpha}\partial X_{2\beta}} = - \eta^{\alpha \beta}, \ X_1.X_2 = -(1+u) \, .
\end{equation}
Substituting those relations in \eqref{eefourth1}, we have
\begin{equation} \label{eefourth3}
\begin{split}
\mathscr{I}_4 &= -\frac{4 \pi^3 (1+u)}{\nu^2(1+\nu^2)} W_{1\alpha}W_{2\alpha}  \left[ - \eta^{\alpha \beta} \partial_u \Omega(u, \nu) + X_2^{\alpha} X_1^{\beta} \ \partial^2_u \Omega(u, \nu)  \right] \\ &= -\frac{4 \pi^3 (1+u)}{\nu^2(1+\nu^2)} \left[ - (W_1.W_2)\partial_u \Omega(u, \nu) + (W_1.X_2)(W_2.X_1) \ \partial^2_u \Omega(u, \nu)  \right] \, .
\end{split}
\end{equation}
Therefore, the total integral $\mathscr{I}$ is given by
\begin{equation} \label{eefourth4}
\begin{split}
\mathscr{I} & = \frac{1+\nu^2}{4 \pi^3}(\mathscr{I}_1+\mathscr{I}_2+\mathscr{I}_3+\mathscr{I}_4) \\ &= \frac{1}{\nu^2} \Bigg( (W_1.W_2) \Big( \Omega(u, \nu) (1+\nu^2) + (1+u) \partial_u \Omega(u,\nu) \Big) + (W_2.X_1)(X_2.W_1) \\& ~~~ \Big( -2 \partial_u \Omega(u, \nu) - (1+u) \partial^2_u \Omega(u, \nu) \Big) \Bigg) \, .
\end{split}
\end{equation}
The functions $\Omega_1$ and $\Omega_2$ can be also written in terms of the scalar harmonics as shown in \eqref{Omegrel}. Using these relation, the integral reduces to
\begin{equation}
\begin{split}
\mathscr{I} = (W_1.W_2) \Omega_1(u, \nu) + (W_2. X_1) (W_1. X_2) \Omega_2(u, \nu) = \Omega^{(e)}(u, \nu).
\end{split}
\end{equation}
%

\subsection{Parity-odd harmonics: odd-even split}

The even spin-1 bulk-boundary propagator in $d = 2$ is given in \eqref{spin1bulkbdy} while the odd spin-1 bulk-boundary propagator in $d = 2$ is
\begin{equation} \label{oddbulkbdyprop}
\widetilde \Pi^{(o)}(X, P; W, Z; \Delta) = C_{\Delta}\frac{\epsilon_{ABCD} X^{A}P^{B}W^{C}Z^{D}}{(-2 X.P)^{\Delta + 1}}, \qquad C_{\Delta} = -\frac{\Delta}{\pi} \, ,
\end{equation}
where $C_{\Delta}$ is a normalisation constant. We are interested in evaluating the following integral
\begin{equation} \label{reqdintoe}
\mathscr{I} = \frac{\nu^2}{\pi} \ \int_{\partial} d P \ \widetilde \Pi^{(o)}(X_1, P; W_1, K^{(b)}_Z; 1+ i \nu) \widetilde \Pi^{(e)}(P, X_2; Z, W_2; 1- i \nu) \, .
\end{equation}
Substituting the expressions, we get
\begin{equation} \label{reqdintoe2}
\begin{split}
& \mathscr{I} =  \frac{\nu^2(1+\nu^2)}{2 i \nu \pi^3} \int_{\partial} d P \frac{\epsilon_{ABCD} X_1^{A}P^{B}W_1^{C}\frac{\partial }{\partial Z_D}}{(-2 X_1.P)^{2+i\nu}} \frac{ (-2X_2 . P)(W_2.Z) + 2 (W_2.P) (Z.X_2)  }{(-2 X_2.P)^{2-i \nu}} \\ & = \frac{\nu^2(1+\nu^2)}{2 i \nu \pi^3} \int_{\partial} d P \frac{\epsilon_{ABCD} X_1^{A}P^{B}W_1^{C}}{(-2 X_1.P)^{2+i \nu}(-2 X_2.P)^{2-i \nu}} \Bigg( (-2 X_2.P) W_2^D + 2 (W_2.P) X_2^D  \Bigg) \, .
\end{split}
\end{equation}
The integral of the term
\begin{equation} \label{reqdintoefirstTerm}
\begin{split}
\mathscr{I}_1 & = \ \int_{\partial} d P \frac{\epsilon_{ABCD} X_1^{A}P^{B}W_1^{C} W_2^{D}}{(-2 X_1.P)^{2+i\nu}(-2 X_2.P)^{1-i\nu}} =  \int_{\partial} d P \frac{\epsilon_{ABCD} X_1^{A}P^{B}W_1^{C} W_2^{D}}{(-2 X_1.P)^{2+i\nu}(-2 X_2.P)^{1-i\nu}} \\ &= \frac{\epsilon_{ABCD} X_1^{A}W_1^{C} W_2^{D}}{ 2(1+i \nu)} \frac{\partial }{\partial X_{1B} }\int_{\partial} d P \frac{1}{(-2 X_1.P)^{2+i\nu}(-2 X_2.P)^{1-i\nu}} \\ & = \frac{4 \pi^3}{\nu^2} \frac{\epsilon_{ABCD} X_1^{A}W_1^{C} W_2^{D}}{ 2(1+i \nu)}  \frac{\partial u}{\partial X_{1B}} \partial_u \Omega(u, \nu) \\& = - \frac{2 \pi^3}{\nu^2} \frac{1}{(1+i \nu)} \epsilon_{ABCD} X_1^{A} X_2^{B} W_1^{C} W_2^{D} \partial_u \Omega(u, \nu) \, ,
\end{split}
\end{equation}
where we have used \eqref{eefourth2} to get to the last equality in \eqref{reqdintoefirstTerm}. The remaining integral which is the term
\begin{equation} \label{reqdintoesecTerm}
\begin{split}
& \mathscr{I}_2 = \ 2 \int_{\partial} d P  \frac{\epsilon_{ABCD} X_1^{A}P^{B}W_1^{C} X_2^{D} (W_2. P)}{(-2 X_1.P)^{2+i\nu}(-2 X_2.P)^{2-i\nu}} \\& =  \ 2 \ \epsilon_{ABCD} X_1^{A} W_1^{C} X_2^{D}  W_{2 M}  \int_{\partial} d P \frac{ P^{B}  P^{M}}{(-2 X_1.P)^{2+i\nu}(-2 X_2.P)^{2-i\nu}} \\& = \ 2 \ \frac{\epsilon_{ABCD} X_1^{A} W_1^{C} X_2^{D}  W_{2 M}}{4 (1+\nu^2)}  \frac{\partial}{\partial X_{1B}} \frac{\partial}{\partial X_{2M}} \int_{\partial} d P \frac{1}{(-2 X_1.P)^{1+i\nu}(-2 X_2.P)^{1-i\nu}} \\&= \  \frac{4 \pi^3}{\nu^2} \frac{\epsilon_{ABCD} X_1^{A} W_1^{C} X_2^{D}  W_{2 M}}{2 (1+\nu^2)}  \frac{\partial}{\partial X_{1B}} \frac{\partial}{\partial X_{2M}}  \Omega(u, \nu) \\&= \  \frac{4 \pi^3}{\nu^2} \frac{\epsilon_{ABCD} X_1^{A} W_1^{C} X_2^{D}  W_{2 M}}{2 (1+\nu^2)}   \left[ \frac{\partial^2 u}{\partial X_{1B} \partial X_{2M}} \partial_u \Omega(u, \nu) + \frac{\partial u}{\partial X_{1B}} \frac{\partial u}{ \partial X_{2M}} \partial^2_u \Omega(u, \nu)  \right]  \\&= -  \frac{4 \pi^3}{\nu^2} \frac{\epsilon_{ABCD} X_1^{A} W_1^{C} X_2^{D}  W_{2 M}}{2 (1+\nu^2)}  \eta^{BM} \partial_u \Omega(u, \nu) \\& = - \frac{2 \pi^3}{\nu^2} \frac{\epsilon_{ABCD} X_1^{A} W_1^{C} X_2^{D}  W_{2}^B}{(1+\nu^2)} \partial_u \Omega(u, \nu) \, ,
\end{split}
\end{equation}
where we have used the anti-symmetry of the $\epsilon$ tensor and \eqref{eefourth2} to get to the last equality in \eqref{reqdintoesecTerm}. Combining $\mathscr I_1$ and $\mathscr I_2$, we have
\begin{equation} \label{reqdintoe3}
\begin{split}
\mathscr{I} & =  \frac{\nu^2(1+\nu^2)}{2 i \nu \pi^3} (\mathscr{I}_1+\mathscr{I}_2) \\&= \frac{1+\nu^2}{i \nu}  \Bigg( - \frac{\epsilon_{ABCD} X_1^{A} X_2^{B} W_1^{C} W_2^{D}}{(1+i \nu)}  - \frac{\epsilon_{ABCD} X_1^{A} W_1^{C} X_2^{D}  W_{2}^B}{(1+\nu^2)}  \Bigg) \partial_u \Omega(u, \nu) \\& = \frac{1+\nu^2}{i \nu} \epsilon_{ABCD} X_1^{A} X_2^{B} W_1^{C} W_2^{D}  \Bigg( \frac{1}{(1+\nu^2)} - \frac{1}{(1+i \nu)} \Bigg) \partial_u \Omega(u, \nu) \\& = \epsilon_{ABCD} X_1^{A} X_2^{B} W_1^{C} W_2^{D} \partial_u \Omega(u, \nu) \, .
\end{split}
\end{equation}
Using \eqref{O3Orel}, we have
$\mathscr{I}= \Omega^{(o)}(u, \nu)$.

\subsection{Parity-even harmonics: odd-odd split}
%
We use the odd spin-1 bulk-boundary propagator as given in \eqref{oddbulkbdyprop} to evaluate the integral
\begin{equation} \label{reqdinoo1}
\begin{split}
\mathscr{I} & = \frac{\nu^2}{\pi} \ \int_{\partial} d P \ \widetilde \Pi^{(o)}(X_1, P; W_1, K^{(b)}_Z; 1+ i \nu) \widetilde \Pi^{(o)}(P, X_2; Z, W_2; 1- i \nu) \\
& = \frac{\nu^2(1+\nu^2)}{\pi^3} \ \int_{\partial} d P \frac{\epsilon_{ABCD} X_1^{A}P^{B}W_1^{C}\frac{\partial }{\partial Z_D}}{(-2 X_1.P)^{2+i\nu}} \  \frac{\epsilon_{EFGH} P^{E}X_2^{F}Z^{G}W_2^{H}}{(-2 X_2.P)^{2-i\nu}} \\ & = \frac{\nu^2(1+\nu^2)}{\pi^3} \epsilon_{ABCD} \epsilon_{EFGH}  \eta^{DG} X_1^{A} W_1^{C}  X_2^{F} W_2^{H} \int_{\partial} d P \frac{P^{B}}{(-2 X_1.P)^{2+i\nu}}   \frac{ P^{E}}{(-2 X_2.P)^{2-i\nu}} \, .
\end{split}
\end{equation}
The term inside the integral can be written as
\begin{equation}
\frac{P^{B}}{(-2 X_1.P)^{2+i\nu}} \  \frac{ P^{E}}{(-2 X_2.P)^{2-i\nu}} = \frac{1}{4(1+ \nu^2)} \frac{\partial}{\partial X_{1B}} \frac{\partial}{\partial X_{2E}} \frac{1}{(-2 X_1.P)^{1+i\nu} (-2 X_2.P)^{1-i\nu}} \, .
\end{equation}
Thus, equation \eqref{reqdinoo1} becomes
\begin{align} \nonumber
 & \frac{\nu^2}{4 \pi^3} \epsilon_{ABCD} \epsilon_{EFGH} \ \eta^{DG} X_1^{A} W_1^{C}  X_2^{F} W_2^{H} \    \frac{\partial^2}{\partial X_{1B} \partial X_{2E}} \int_{\partial} d P  \frac{1}{(-2 X_1.P)^{1+i\nu} (-2 X_2.P)^{1-i\nu}} \\ \label{reqdintoo2} &= \epsilon_{ABCD} \ \epsilon_{EFGH} \ \eta^{DG} X_1^{A} W_1^{C}  X_2^{F} W_2^{H}  \frac{\partial}{\partial X_{1B}} \frac{\partial}{\partial X_{2E}} \Omega(u, \nu) \\ &  \nonumber = \epsilon_{ABCD} \epsilon_{EFGH} \eta^{DG} X_1^{A} W_1^{C}  X_2^{F} W_2^{H} \left(\frac{\partial^2 u}{\partial X_{1B}\partial X_{2E}} \partial_u \Omega(u, \nu) + \frac{\partial u}{\partial X_{2E}} \frac{\partial u}{\partial X_{1B}} \partial^2_u \Omega(u, \nu) \right) \, ,
\end{align}
where we used \eqref{scalarsplit} to write the second equality in \eqref{reqdintoo2}. Substituting the relations obtained in \eqref{eefourth2}, we have
\begin{equation} \label{reqdintoo3}
\begin{split}
 \mathscr{I} = \epsilon_{ABCD} \ \epsilon_{EFGH} \ \eta^{DG} X_1^{A} W_1^{C}  X_2^{F} W_2^{H} \ \left(- \eta^{BE} \partial_u \Omega(u, \nu) + X_1^E X_2^B \  \partial^2_u \Omega(u, \nu) \right)
\end{split}
\end{equation}
Using the identities of product of two Levi-Civita symbols, the transverse conditions $X_1. W_1 = X_2. W_2 = 0$ and the $AdS$ equation $(X_1)^2 = (X_2)^2 = -1$, we find
\begin{equation}
\begin{split}
 & \epsilon_{ABCD} \ \epsilon_{EFGH} \ \eta^{DG} \eta^{BE} X_1^{A} W_1^{C}  X_2^{F} W_2^{H} = -2 ( (X_1.X_2) (W_1.W_2)-(X_1.W_2)(W_1.X_2))\\
 & \epsilon_{ABCD} \ \epsilon_{EFGH} \ \eta^{DG} X_1^{A} W_1^{C}  X_2^{F} W_2^{H} X_1^{E} X_2^{B} = (X_1.W_2)(W_1.X_2)(X_1.X_2)) \\ & ~~~~~~~~~~~~~~~~~~~~~~~~~~~~~~~~~~~~~~~~~~~~~~~~~~~~~~~~~~~~~ -((X_1.X_2)^2-1) (W_1.W_2) ,
\end{split}
\end{equation}
which after substitution in \eqref{reqdintoo3} and a suitable rearrangement gives
\begin{equation} \label{reqdintoo5}
\begin{split}
& \mathscr{I} =   -( [ ((X_1.X_2)^2-1) \partial^2_u \Omega(u, \nu) -2 (X_1.X_2) \partial_u \Omega(u, \nu) \ ] (W_1.W_2) \\ & ~~~~~~~~~~~~~~~ +[2 \ \partial_u \Omega(u, \nu) - (X_1.X_2) \ \partial^2_u \Omega(u, \nu) ] (X_1.W_2)(X_2.W_1) \ ) \, .
\end{split}
\end{equation}
In terms of the chordal distance $u$, the above expression becomes
\begin{equation} \label{reqdintoo6}
\begin{split}
\mathscr{I} = - & \  [ u(u+2) \ \partial^2_u \Omega(u, \nu) + 2 (1+u) \ \partial_u \Omega(u, \nu) \ ] (W_1.W_2) \\ & ~~~~~~ - [2 \ \partial_u \Omega(u, \nu) +(1+u) \ \partial^2_u \Omega(u, \nu) ] (X_1.W_2)(X_2.W_1) \, .
\end{split}
\end{equation}
Finally using \eqref{Omegrel}, we have
\begin{equation} \label{reqdintoo7}
\begin{split}
  \mathscr{I} = \left( (W_1.W_2) \Omega_1(u, \nu)+ \Omega_2(u, \nu) (X_1.W_2)(X_2.W_1) \right) = \nu^2 \Omega^{(e)}(u, \nu) \, .
\end{split}
\end{equation}
This completes our proof of the split representations.

\section{Evaluation of integrals appearing in the propagators} \label{app:int}

We evaluate the even part of the integrals appearing in the Proca theory \eqref{fulpropproca} and  massive CS theory (\eqref{mcsevintdef}, \eqref{mcsoa}) by writing it in terms of the scalar harmonics using identities \eqref{Omegrel}. Then, we use the split representation of the scalar harmonic functions \eqref{scalsplit} to evaluate those integrals. The details of this evaluation has been provided in the following subsections.

\subsection{Integrals in the Proca theory} \label{Assec:proca}
%
In this subsection, we shall evaluate the integrals appearing in \eqref{fulpropproca}. At first, we begin with the integral over
the parity-even harmonic functions
\begin{equation} \label{app:proca1}
\begin{split}
\mathcal I_1 &= \int d \nu \frac{\Omega_{\alpha \beta}^{(e)}(x,y;\nu)}{\nu^2 + (\Delta - 1)^2} \\ &= - \frac{\partial^2 u}{\partial x^{\alpha} \partial y^{\beta}} \mathbb D^{(1)} \left(\mathcal F^{(e)}(u)\right) + \frac{\partial u}{\partial x^{\alpha} }  \frac{\partial u}{\partial y^{\beta} }\mathbb D^{(2)} \left(\mathcal F^{(e)}(u)\right) \, ,
\end{split}
\end{equation}
where
\begin{equation} \label{app:proca2}
\begin{split}
\mathcal F^{(e)}(u) &= \frac{i}{2\pi} \int_{-\infty}^{\infty} d \nu \ \frac{\Pi (u,1+i \nu )-\Pi (u,1-i \nu )}{\nu \left(\nu^2 + (\Delta - 1)^2 \right)} \\
& = \frac{i}{8 \pi ^2 \sqrt{u (u+2)}}  \int_{-\infty}^{\infty} d \nu \ \left(\frac{ e^{-2 i A \nu} }{\nu \left(\nu^2 + (\Delta - 1)^2 \right)} -  \ \frac{ e^{2 i A \nu} }{\nu \left(\nu^2 + (\Delta - 1)^2 \right)} \right) \, ,
\end{split}
\end{equation}
with $A = \ln \left(\frac{\sqrt{u}+\sqrt{u+2}}{\sqrt{2}}\right)$, which is always non-negative since $u \geq 0$. We observe that the exponential $e^{-2 i A \nu}$ in \eqref{app:proca2}  goes to $0$ as $\text{Im}(\nu) \to -\infty $. Similarly, the exponential $e^{2 i A \nu} \to 0$ as $\text{Im}(\nu) \to \infty$. Thus, we perform the integral in complex $\nu$-plane via choosing distinct contours for the two exponentials.

Note that both the terms in this integral contains three simple poles at $$\nu = 0, \pm \ i (\Delta - 1).$$ To regulate the pole at $\nu = 0$, we use a pole prescription where the pole is pushed along the imaginary axis to the upper half plane, i.e. the pole is shifted by $+ i \epsilon$
\footnote{We have verified that the alternative pole prescription of shifting the pole by $-i \epsilon$, pushing it into the lower half plane, also gives the same final
result.}. We close the contour for the part containing the exponential $e^{-2 i A \nu}$ in the lower-half plane thus picking up the pole at $\nu = - \ i (\Delta - 1)$. While for the term containing $e^{2 i A \nu}$, we close the contour in the upper-half plane, thus picking up contributions from the poles
$\nu = i \epsilon, ~ i (\Delta - 1)$. The final result after putting together contribution from all these poles, is given by
\begin{equation}
\mathcal F^{(e)}(u) = \frac{\left(\sqrt{u}+\sqrt{u+2}\right)^{-2 \Delta } \left(\left(\sqrt{u}+\sqrt{u+2}\right)^{2 \Delta }-2^{\Delta } (u+1)-2^{\Delta } \sqrt{u(u+2)}\right)}{4 \pi  (\Delta -1)^2 \sqrt{u (u+2)}} \, .
\end{equation}
Next, let us consider the longitudinal part of the integral in \eqref{fulpropproca}
\begin{equation}\label{app:intprocaLong}
\begin{split}
\mathcal I_2 &= \int d\nu \frac{\Sigma_{\alpha \beta} (x,y;\nu)}{(\nu^2+1)(\Delta-1)^2} \\ &= \frac{\partial^2 u}{\partial x^{\alpha} \partial y^{\beta}} \partial_u \left(\mathcal F^{(l)}(u)\right) + \frac{\partial u}{\partial x^{\alpha} }  \frac{\partial u}{\partial y^{\beta} } \partial_u^2 \left(\mathcal F^{(l)}(u)\right) \, ,
\end{split}
\end{equation}
where
\begin{equation} \label{app:intprocaLong2}
\begin{split}
\mathcal F^{(l)}(u) &= \int d \nu \frac{\Omega(x,y;\nu)}{(\nu^2+1)(\Delta-1)^2}
\\& = \frac{i}{ 8 \pi ^2 (\Delta-1)^2 \sqrt{u (u+2)} } \left( \int_{-\infty}^{\infty} d \nu \ \frac{\nu \ e^{-2 i A \nu} }{\nu^2+1} - \int_{-\infty}^{\infty} d \nu \ \frac{ \nu \ e^{2 i A \nu} }{\nu^2+1} \right) \, .
\end{split}
\end{equation}
The poles of this integral are at $\nu = \pm i$ and no pole prescriptions are necessary here. As done previously, closing the contour in the
upper and lower-half planes respectively for the two terms, we pick up contributions from the poles by evaluating the residues appropriately.
We obtain the following final result
\begin{equation}
\mathcal F^{(l)}(u) = \frac{1}{2 \pi  (\Delta-1)^2 \sqrt{u (u+2)} \left(\sqrt{u}+\sqrt{u+2}\right)^2} \, .
\end{equation}
Thus, the full propagator $\Pi^{(e)}_{\alpha \beta}(x,y; \Delta)$ in \eqref{fulpropproca} is given by
\begin{equation} \label{app:procafull1}
\begin{split}
\Pi_{\alpha \beta}(x,y; \Delta) = - \frac{\partial^2 u}{\partial x^{\alpha} \partial y^{\beta}} &\left( \mathbb D^{(1)}  \mathcal F^{(e)}(u) - \partial_u  \mathcal F^{(l)}(u)  \right) \\ &+ \frac{\partial u}{\partial x^{\alpha} }  \frac{\partial u}{\partial y^{\beta} } \left( \mathbb D^{(2)} \mathcal F^{(e)}(u) + \partial_u^2 \mathcal F^{(l)}(u) \right) \, .
\end{split}
\end{equation}
We have checked that $\left( \mathbb D^{(1)}  \mathcal F^{(e)}(u) - \partial_u  \mathcal F^{(l)}(u)  \right) = g_0(u)$ and $ \left( \mathbb D^{(2)} \mathcal F^{(e)}(u) + \partial_u^2 \mathcal F^{(l)}(u) \right) = g_1(u)$, where $g_{0}$ and $g_1$ are defined in \eqref{g0g1def}.

Note that these integrals has also been reported in the appendix of \cite{Ankur:2023lum}, where the parity-even split representation of the spin-1 vector harmonics \eqref{evOpi} was used directly. Here we have reconfirmed the expected results by using the split representation of the scalar harmonics instead of \eqref{evOpi}.

\subsection{Integrals in the Massive CS theory}\label{Assec:MCS}
%
In this subsection, we evaluate the integrals appearing in \eqref{mcsprints}. These integrals are very similar to those appearing in \eqref{fulpropproca} and the computation proceeds parallel to \ref{Assec:proca}. The transverse parity-even part of \eqref{mcsprints} is given by \eqref{mcsevintdef}, which involves
the following integral over the scalar harmonics
\begin{equation} \label{app:intmasscs}
\begin{split}
\mathcal F^{(e)}(u) &= \int d\nu \frac{\Omega(u)}{\nu^2 (\nu - i m )} =  \frac{i}{2\pi} \int_{-\infty}^{\infty} d \nu \ \frac{\Pi (u,1+i \nu )-\Pi (u,1-i \nu )}{\nu (\nu - i m)} \\
& = \frac{i}{2^3 \left(\pi ^2 \sqrt{u (u+2)}\right)} \left( \int_{-\infty}^{\infty} d \nu \ \frac{ e^{-2 i A \nu} }{\nu (\nu - i m)} - \int_{-\infty}^{\infty} d \nu \ \frac{ e^{2 i A \nu} }{\nu (\nu - i m)} \right) \, .
\end{split}
\end{equation}
Here, we have again used the split representation of the scalar harmonics \eqref{scalsplit}.
Note that just like \eqref{app:proca2},  $A = \ln \left(\frac{\sqrt{u}+\sqrt{u+2}}{\sqrt{2}}\right) \geq 0$ and consequently the exponentials
in the two integrals lead to a choice of closing of contour similar to \eqref{app:proca2}. However, in this case, we have two simple poles at $\nu = 0, i m$. Recall that $m$ can be positive or negative, depending on the sign of $\kappa$. The
$\nu = 0$ pole is shifted by $+i \epsilon$. Again, we have checked that a choice of $-i \epsilon$ instead, does not affect the final answer. Subsequently,
the contour integrals are performed by collecting the residues over these poles and we get
\begin{equation}
\mathcal F^{(e)}(u) = \frac{i \ \text{sgn}(m) \left(1-2^{|m|} \left(\sqrt{u}+\sqrt{u+2}\right)^{-2 |m|}\right)}{4 \pi  |m| \sqrt{u (u+2)}} \, .
\end{equation}
Similarly, the parity-odd integral \eqref{mcsoa} is performed as follows
\begin{equation} \label{app:intmasscsOdd}
\begin{split}
\mathcal F^{(o)}(u) & =  \int_{-\infty}^{\infty} d \nu \ \frac{\Omega_3(u)}{\nu (\nu-i m)} \\ & = \frac{1}{8 \pi ^2 (u (u+2))^{3/2} } \Bigg ( \int_{-\infty}^{\infty} d \nu \ \frac{\left(\nu  \sqrt{u (u+2)} -i (u+1) \right) e^{-2 i A \nu}}{(\nu -i m)} \\ & + \int_{-\infty}^{\infty} d \nu \ \frac{\left(\nu  \sqrt{u (u+2)} + i (u+1) \right) e^{2 i A \nu}}{(\nu -i m)}
\Bigg) \, .
\end{split}
\end{equation}
Here, it is straightforward to use the expression for $\Omega_3$ reported in \eqref{O3def}, instead of expressing it in terms of scalar harmonics.
While using \eqref{O3def}, we have converted the $\sin$ and $\cos$ functions into their exponential counterparts. Converting the $\sinh^{-1}(x)$ into $\ln (x+ \sqrt{1+x^2})$,
we find that even in \eqref{app:intmasscsOdd}, $A$ is given by the same expression $$A=\ln \left(\frac{\sqrt{u}+\sqrt{u+2}}{\sqrt{2}}\right) \geq 0.$$ Hence,
closing the contours similar to \eqref{app:proca2} and \eqref{app:intmasscs}, we pick up the residue at the $\nu = i m$ pole. For any sign of $m$, we obtain the following result
\begin{equation}
\mathcal F^{(o)}(u) = -\frac{2^{|m|} \left(\sqrt{u}+\sqrt{u+2}\right)^{-2 |m|} \left(|m| \sqrt{u (u+2)}+u+1\right)}{4 \pi  (u(u+2))^{3/2}} \, .
\end{equation}
Finally, the integral appearing in the longitudinal term \eqref{mcsLintdef} is performed as follows
\begin{equation} \label{app:intmasscsLong}
\begin{split}
\mathcal F^{(l)}(u) & = \int d\nu \frac{\Omega(u)}{(\nu^2 + 1)} = \frac{i}{ 8 \pi ^2 \sqrt{u (u+2)} } \left( \int_{-\infty}^{\infty} d \nu \ \frac{\nu \ e^{-2 i A \nu} }{\nu^2+1} - \int_{-\infty}^{\infty} d \nu \ \frac{ \nu \ e^{2 i A \nu} }{\nu^2+1} \right) \, .
\end{split}
\end{equation}
For this integral, we have again used the split representation of the scalar harmonics.
Here, the poles are at $\nu = \pm i$ avoiding the real-axis. Hence no pole prescription is necessary and we can perform the integrals similar to \eqref{app:intprocaLong2}.   From the residues at $\nu = \pm i$, we obtain the following result
\begin{equation}
\mathcal F^{(l)}(u) = \frac{1}{2 \pi  \sqrt{u (u+2)} \left(\sqrt{u} + \sqrt{u+2}\right)^2} \, .
\end{equation}


\section{Conventions for SO(3,1) generators} \label{app:gens}
%
For our choice of Poincare patch coordinates $\{z,x,y\}$ in \eqref{met}, the killing vectors fields corresponding to $SO(3,1)$ isometry of AdS$_3$ are given by
\begin{equation}\label{so31gen}
 \begin{split}
  & \text{Dilation} : \xi_D = z \partial_z + x \partial_x + y \partial_y ,\\
  & \text{Rotation} : \xi_R = - y \partial_x + x \partial_y \\
  & \text{SCT (along x):} ~ \xi_{K_x} = - 2 x z \,\partial_z - \left( x^2 - y^2  - z^2 \right) \partial_x - 2 x y \, \partial_y \\
  & \text{SCT (along y):} ~ \xi_{K_y} = - 2 y z \, \partial_z - 2 x y \, \partial_x + \left( x^2 - y^2  + z^2 \right)  \partial_y \\
  & \text{Translation (along x):} ~ \xi_{P_{x}} =  \partial_x \, ,\\
  & \text{Translation (along y):} ~ \xi_{P_y} =  \partial_y \, ,\\
 \end{split}
\end{equation}
These generators map to the  $SU(2) \times SU(2)$ generators, which are enumerated below in terms of the familiar holomorphic and anti-holomorphic notation of 2D CFTs
\begin{equation}\label{su2su2gen}
 \begin{split}
  L_0 & = - \frac{1}{2} \left( D - i \, R\right) \, , \\
  L_{-1} &= - \frac{1}{2} \left( P_x - i \, P_y\right) \, , \\
  L_{1} &= - \frac{1}{2} \left( K_x + i \, K_y\right) \, ,
 \end{split}
 \quad \quad \quad \quad
 \begin{split}
  \bar{L}_0 & = - \frac{1}{2} \left( D + i \, R\right) \, ,  \\
  \bar{L}_{-1} &= - \frac{1}{2} \left( P_x + i \, P_y\right) \, , \\
  \bar{L}_{1} &= - \frac{1}{2} \left( K_x - i \, K_y\right) \, .
 \end{split}
\end{equation}
The two quadratic Casimirs can be easily defined in terms of the generators in \eqref{su2su2gen} as follows
\begin{equation}\label{casimir}
 \begin{split}
  C &= L_0^2 - \frac{1}{2} \left( L_{-1} L_1 + L_1 L_{-1} \right) \, , \\
  \bar{C} &= \bar {L}_0^2 - \frac{1}{2} \left( \bar{L}_{-1} \bar{L}_1 + \bar{L}_1 \bar{L}_{-1} \right) \, .
 \end{split}
\end{equation}
In this complexified notation, a representation is labelled by $(h, \bar{h})$, which are the eigenvalues of the $L_0$ and $\bar{L}_0$ for the primaries of an infinite dimensional unitary representation. Over such a representation, the Casimirs $C$, $\bar{C}$ evaluates to $h(h-1)$ and
$\bar {h}(\bar{h} - 1)$ respectively. Using the relation \eqref{su2su2gen}, these Casimirs can be expressed in terms of Lie derivatives with respect to the isometries \eqref{so31gen}. Further, all the generators in \eqref{so31gen}, and hence the Casimirs acts on our harmonic functions through these Lie derivatives. Thus, the $(h, \bar{h})$ representation labels of our harmonic functions can be read off from the action of these Casimirs.

\bibliographystyle{JHEP}
\bibliography{CSinAdS}
\end{document}